\documentclass[a4paper,UKenglish,cleveref, autoref]{lipics-v2021}

\title{Space-Optimal Spanning Tree Algorithms in Bounded-Deletion Streams}

\author{Tsun-Ming Cheung}{Department of Computer Science, National Tsing Hua University}{bencheung@cs.nthu.edu.tw}{https://orcid.org/0009-0003-7698-5088}{}
\author{Yu Hsiang}{Department of Computer Science and Information Engineering, National Taiwan University.}{kasperhsiang@gmail.com}{https://orcid.org/0009-0001-8471-9375}{}
\author{Shang-En Huang}{Department of Computer Science and Information Engineering, National Taiwan University.}{sehuang@csie.ntu.edu.tw}{}{}
\author{Peng-Ting Lin}{Institute of Information Science, Academia Sinica , Taiwan}{ptlin@iis.sinica.edu.tw}{https://orcid.org/0009-0009-0776-9318}{}
\author{Meng-Tsung Tsai}{Institute of Information Science, Academia Sinica, Taiwan}{mttsai@iis.sinica.edu.tw}{https://orcid.org/0000-0002-2243-8666}{}

\authorrunning{T.-M. Cheung, Y. Hsiang, S.-E. Huang, P.-T. Lin, M.-T. Tsai} 

\Copyright{Tsun-Ming Cheung and Yu Hsiang and Shang-En Huang and Peng-Ting Lin and Meng-Tsung Tsai} 



\category{} 

\relatedversion{} 

\hideLIPIcs

\nolinenumbers 

\EventEditors{}
\EventNoEds{0}
\EventLongTitle{}
\EventShortTitle{STACS}
\EventAcronym{STACS}
\EventYear{}
\EventDate{}
\EventLocation{}
\EventLogo{}
\SeriesVolume{}
\ArticleNo{1}

\usepackage{hyperref}

\hypersetup{colorlinks=true,
 linkcolor=red,
citecolor=blue,
urlcolor=magenta
}

\usepackage{booktabs}
\usepackage{multicol}
\usepackage{amsthm}
\usepackage{thmtools}
\usepackage{mathtools}
\usepackage{todonotes}
\usepackage[ruled,vlined,linesnumbered]{algorithm2e}

\def\INDEX{\textup{\textsf{INDEX}}}
\def\UR{\textup{\textsf{UR}}^{\subset}}

\newcommand{\abs}[1]{|#1|}
\newcommand{\floor}[1]{\lfloor#1\rfloor}

\newcommand{\floorB}[1]{\left\lfloor#1\right\rfloor}

\newcommand{\set}[1]{\{#1\}}
\newcommand{\setB}[1]{\left\{#1\right\}}
\def\defeq{\coloneqq}

\def\N{\mathbb{N}}
\def\R{\mathbb{R}}
\def\Dcc{\textup{\textsf{D}}^\to}
\def\Rcc{\textup{\textsf{R}}^\to}
\def\calX{\mathcal{X}}
\def\calY{\mathcal{Y}}
\def\calZ{\mathcal{Z}}
\def\scW{\mathscr{W}}
\def\scA{\mathscr{A}}
\def\tdB{\tilde{B}}

\def\I{\mathbf{I}}
\def\H{\mathbf{H}}
\def\L{\mathbf{L}}
\def\KL{\mathbf{D}_{\mathrm{KL}}}
\def\ent{\textup{h}_2}
\def\TV{\mathbf{d}_{\mathrm{TV}}}
\DeclareMathOperator*{\E}{\mathbb{E}}
\def\Msg{\textup{\textsf{Msg}}}
\def\Dec{\textup{\textsf{Dec}}}

\newcommand{\doublefrac}[2]{%
  \begin{array}{@{}c@{}}
    #1 \\[0ex] \hline
    \noalign{\vskip 1pt}\hline
    #2
  \end{array}%
}
\newcommand{\CRE}[3]{
\KL\left(\doublefrac{#1}{#2} \;\middle|\; #3\right)
}

\DeclareMathOperator{\id}{id}
\DeclareMathOperator{\supp}{supp}

\definecolor{purple}{HTML}{1A1AB3}

\DeclareEmphSequence{\color{purple}\itshape}

\AtBeginEnvironment{algorithm}{%
  \DeclareEmphSequence{\itshape}
}
\AtEndEnvironment{algorithm}{%
  \DeclareEmphSequence{\color{purple}\itshape}
}

\AtBeginEnvironment{thebibliography}{%
  \DeclareEmphSequence{\itshape}
}
\AtEndEnvironment{thebibliography}{%
  \DeclareEmphSequence{\color{purple}\itshape}
}
\crefname{appendix}{Appendix}{Appendices}
\Crefname{appendix}{Appendix}{Appendices}

\makeatletter
\@ifundefined{newcounteralias}{}{%
  \renewcommand\thmt@autorefsetup{%
    \@xa\def\csname\thmt@envname autorefname\@xa\endcsname
      \@xa{\thmt@thmname}%
  }%
}
\makeatother

\newtheorem{fact}[theorem]{Fact}
\crefname{fact}{Fact}{Facts}
\Crefname{fact}{Fact}{Facts}
\def\1{\mathbf{1}}

\begin{document}

\maketitle

\begin{abstract}

We determine the optimal single-pass space complexity of computing a spanning forest of an $n$-vertex simple graph in a stream of edge insertions and at most $k$ deletions. For $1\le k\le n^2$ and any fixed failure probability $0<\delta\le 1/3$, the optimal randomized space is $\Theta_\delta(n\log n\,\log^2(2k))$ bits. For every fixed $c>0$, our upper and lower bounds match throughout the range $n^{-c}\le\delta\le 1/3$. At $\delta=n^{-c}$, the optimal space is $\Theta_c\!\left(n\log^2 n\,\log\!\left(1+\frac{k}{\log n}\right)\right)$ bits.

We also give a dynamic data structure with polylogarithmic worst-case update time and $O(\log n/\log\log n)$ worst-case connectivity query time. For every oblivious update sequence with at most $k$ deletions, the probability that it maintains a spanning forest of the current graph after every update is at least $1-n^{-c}$.

For the deterministic setting, we prove that for every $k\ge 1$, the optimal streaming space is $\Theta\!\left(\min\!\left\{n^2, nk\log\!\left(1+\frac{n}{k}\right)\right\}\right)$ bits. The matching upper bound is achieved by maintaining a sparse edge-connectivity certificate.

These results also provide a basis for space-efficient algorithms in the bounded-deletion model for edge- and vertex-connectivity certificates, bipartiteness testing, maximum-leaf spanning-tree sparsification, and approximate minimum spanning trees. These applications build on reductions to spanning-tree or edge-maximal spanning-forest computation, with appropriate control of the deletion bounds and failure probabilities of the resulting subroutines.

\end{abstract}

\clearpage
\begin{nolinenumbers}
\begingroup
\setcounter{tocdepth}{2}
\section*{Contents}
\makeatletter
\@starttoc{toc}
\makeatother
\endgroup
\end{nolinenumbers}

\thispagestyle{empty}
\clearpage
\setcounter{page}{1}

\section{Introduction}

Graph streaming algorithms process a sequence of edge updates using limited memory. In the insertion-only model, edges can only be inserted, whereas the turnstile model allows both insertions and deletions. The bounded-deletion model lies between these two settings: insertions and deletions may be arbitrarily interleaved, but the total number of deletions is bounded by a parameter $k$. This model allows the space complexity to reflect the number of deletions rather than the full generality of turnstile streams. Prior work includes matching-size estimation in sparse graphs under a restricted deletion bound by Cormode, Jowhari, Monemizadeh, and Muthukrishnan~\cite{CormodeHMM17}; maximal matching by Khanna, Konrad, and Dark~\cite{khanna_konrad_dark:ICALP.2025}; and approximate distance queries and spanners by Khanna, Konrad, and Putterman~\cite{KhannaKP26}. In this paper, we study the space complexity of computing a spanning tree in the bounded-deletion model. For disconnected input graphs, our algorithms return an edge-maximal spanning forest consisting of one spanning tree from each connected component. Our lower bounds hold even when the input graph is promised to be connected.

\begin{definition}[Bounded-deletion model]
Let $k\ge 0$ be an integer. In the \emph{bounded-deletion model}, given $n$ and $k$ in advance, an algorithm processes a stream of edge insertions and deletions on a fixed vertex set of size $n$, starting from the empty graph. Each insertion adds a currently absent edge, and each deletion removes a currently present edge, so the graph remains simple throughout the stream. The updates may be arbitrarily interleaved, and an edge may be reinserted after deletion, provided that the total number of deletion operations in the stream does not exceed $k$.
\end{definition}
The insertion-only model corresponds to $k=0$, whereas the turnstile model allows an unbounded number of deletions, with complexity bounds not parameterized by the deletion count. In this paper, we focus on single-pass algorithms, which process the stream sequentially in a single pass.

In insertion-only streams, a spanning tree of an $n$-vertex connected graph can be computed deterministically in a single pass using $O(n\log n)$ bits of space, and the matching $\Omega(n\log n)$ lower bound holds even for randomized algorithms~\cite{SunW15}. In turnstile streams, the optimal single-pass space complexity is $\Theta(n\log^2 n\log(n/\delta))$ bits for failure probability $\delta$. The upper bound follows from the sketching algorithm of Ahn, Guha, and McGregor~\cite{ahn_et_al:SODA.2012}, with its dependence on $\delta$ made explicit by Nelson and Yu~\cite[Appendix A]{nelson_yu:SODA.2019}, who also proved the matching lower bound. We determine the space complexity of spanning-tree algorithms as a function of the deletion bound in the bounded-deletion streaming model. Our main results are summarized in \cref{thm:main}.

\begin{theorem}\label{thm:main}
Let $n\ge 2$ and $1\le k\le n^2$. Consider single-pass randomized algorithms that compute a spanning forest of an $n$-vertex simple graph from a stream containing at most $k$ edge deletions. The optimal space complexity is
\begin{enumerate}
\item $\Theta_\delta\left(n\log n\log^2 k\right)$ bits for every fixed failure probability $0<\delta\le 1/3$;\footnote{Unless otherwise specified, all logarithms are in base two. For asymptotic bounds, we may abbreviate $\log_2(1+x)$ by $\log x$ when $x\ge 1$ throughout the stated parameter range. An expression of the form $\log(1+x)$ is in the ordinary base-two logarithm.}
\item $\Theta_c\!\left(n\log n\,\log(k/\delta)\,\log\!\left(1+\frac{k}{\log(k/\delta)}\right)\right)$ bits for every fixed $c>0$ and every target failure probability $n^{-c}\le\delta\le1/3$.
\end{enumerate}
Both lower bounds hold even when the final graph is promised to be connected.
\end{theorem}
At \(\delta=n^{-c}\), the second bound simplifies to \(\Theta_c\!\left(n\log^2 n\,\log\!\left(1+\frac{k}{\log n}\right)\right)\) bits.
We derive the second bound in \cref{sec:high-prob-upper}. The lower bound for smaller failure probabilities appears in \cref{thm:general_LB}.

Applying a turnstile streaming algorithm directly in the bounded-deletion model uses $\Theta(n\log^3 n)$ bits for either constant or inverse-polynomial failure probability, regardless of the deletion bound. Our bounds quantify the savings obtained by exploiting this bound. For constant failure probability, the space decreases to $\Theta(n\log n\log^2 k)$, giving a factor-$\Theta(\log^2 n/\log^2 k)$ improvement when $k$ is subpolynomial in $n$. For inverse-polynomial failure probability, our bound becomes $\Theta(nk\log n)$ when $1\le k\le\log n$, saving a factor of $\Theta((\log^2 n)/k)$. In particular, both bounds reduce to $\Theta(n\log n)$ for constant $k$, matching the insertion-only complexity, and recover the $\Theta(n\log^3 n)$ turnstile bound when $k=n^{\Theta(1)}$.

\begin{table}[t]
\centering
\label{tab:space-comparison}
\small
\setlength{\tabcolsep}{5pt}
\renewcommand{\arraystretch}{1.25}
\begin{tabular}{@{}llll@{}}
\toprule
Failure probability & Deletion bound & Algorithm space (bits)
  & Improvement factor \\
\midrule
Constant & General $k$
  & $n\log n\,\log^2 k$
  & $\dfrac{\log^2 n}{\log^2 k}$ \\
\addlinespace
Inverse polynomial & General $k$
  & $n\log^2 n\,\log\!\left(1+\dfrac{k}{\log n}\right)$
  & $\dfrac{\log n}{\log(1+k/\log n)}$ \\
\addlinespace
Inverse polynomial & $1\le k\le\log n$
  & $nk\log n$
  & $\dfrac{\log^2 n}{k}$ \\
\midrule
Both regimes & $k=\Theta(1)$
  & $n\log n$ & $\log^2 n$ \\
Both regimes & $k=n^{\Theta(1)}$
  & $n\log^3 n$ & $1$ \\
\bottomrule
\end{tabular}
\caption{Space Complexity in different deletion bound and failure probability regimes.
The turnstile baseline is $\Theta(n\log^2 n \log(n/\delta))$ bits.
All entries in the last two columns are up to $\Theta(\cdot)$;
the improvement factor is the baseline divided by our space bound.
Here $c>0$ is fixed and $1\le k\le n^{2}$, and “Both regimes” refers to constant and inverse-polynomial failure probabilities.}
\end{table}


Beyond the streaming setting, we also maintain an edge-maximal spanning
forest of a dynamic graph with worst-case update guarantees. We build on
the \emph{cutset data structures} of Kapron, King, and
Mountjoy~\cite{KKM13} (see also Gibb et al.~\cite{GKKT15}), which
underlie the first dynamic connectivity algorithms with polylogarithmic
worst-case update time. A cutset structure maintains a forest under
graph-edge updates and forest links and cuts and, given a tree, returns
an edge leaving it with constant probability whenever one exists. In
the nested-forest framework of Kapron, King, and Mountjoy, as refined by
Wang~\cite{Wang15}, the deletion bound ensures that all but at most $2k$
bottom-layer trees already span connected components of the current
graph, so $\Theta(\log k)$ layers suffice instead of
$\Theta(\log n)$. Since only a constant fraction of the
layers must make progress, $\Theta(1+\L(\delta)/\log k)$ cut-set
structures per layer achieve failure probability $\delta$, where $\L(\delta)\defeq \log(1/\delta)$.
 
\begin{theorem}
\label{thm:dynamic}
Fix a constant $c>0$, and let $n\ge2$, $1\le k\le n^2$, and
$n^{-c}\le\delta<1/2$. There is a randomized data structure on a
word RAM with $\Theta(\log n)$-bit words that, given $k$ and $\delta$,
maintains a forest on the vertex set $[n]$ such that,  
for every oblivious sequence of valid updates with at most $k$
deletions,
with probability at least $1-\delta$,
the forest is a spanning forest of the current graph after every update. Each insertion
takes $O_c(L\log^2 n)$ and each deletion
$O_c(L\log^2 n\log (1+k))$ worst-case time, where
$L=\Theta(\log k+\L(\delta))$. Connectivity queries take
$O(\log n/\log\log n)$ worst-case time.
\end{theorem}
 
For $\delta=n^{-c}$, this gives $O(\log^3 n)$ time per insertion and
$O(\log^3 n\log k)$ per deletion, matching the $O(\log^3 n)$
insertion bound of Wang~\cite{Wang15} and improving his $O(\log^4 n)$
deletion bound by a factor $\Theta(\log n/\log k)$, which is
superconstant when $k=n^{o(1)}$. We prove Theorem~\ref{thm:dynamic} in
Appendix~\ref{app:dynamic}.
 
Recently, Meierhans and Probst Gutenberg~\cite{MG26} obtained a Las
Vegas algorithm with expected polylogarithmic worst-case update time,
and Meierhans, Probst Gutenberg, and Yeh~\cite{MPGY26} obtained
polylogarithmic worst-case update time with high probability. Both work against an
adaptive adversary, and their bounds are stated as $\widetilde O(1)$. Our bounds require an oblivious adversary,
but have explicit exponents and adapt to the deletion bound. For
subpolynomial worst-case and amortized bounds, see~\cite{NSW17,CGLNPS20}
and~\cite{HHKPT23}, respectively.

Many space-efficient algorithms in the turnstile model rely on randomization. Notable deterministic exceptions include the multi-pass algorithms for graph coloring of Assadi, Chen, and Sun~\cite{AssadiCS22} and for topological sorting of Chen, Lin, and Tsai~\cite{ChenLT25}. The bounded-deletion promise permits deterministic single-pass algorithms for several graph problems up to moderate deletion bounds, these include maximal matching and approximate maximum matching~\cite{khanna_konrad_dark:ICALP.2025}, as well as approximate distance queries~\cite{KhannaKP26}. Our work gives another space-optimal deterministic single-pass algorithm for computing a spanning tree in the bounded-deletion setting.

\begin{theorem}\label{thm:deterministic}
For every integer $k\ge1$, the deterministic single-pass space
complexity of computing a spanning tree in an $n$-vertex graph
stream with at most $k$ deletions is
\[
 \Theta\left(\min\left\{n^2,\,
 nk\log\left(1+\frac nk\right)\right\}\right)
 \text{ bits}.
\]
The lower
bound holds even for connected final graphs, and the upper
bound also applies to computing an edge-maximal spanning forest
in arbitrary final graphs.
\end{theorem}


\subsection{Technical Overview}
\paragraph*{Bounded-deletion Randomized Algorithm.}
Let \(K=\min\{k,\binom{n}{2}\}\). For \(K\ge1\), our basic algorithm adapts the contraction algorithm of Ahn, Guha, and McGregor (AGM)~\cite{ahn_et_al:SODA.2012} in two ways.
First, we start from the surviving edges of a spanning forest of the \emph{historical graph}, the graph of all edges ever inserted.
Call a component of the current forest \emph{active} if an edge of the final graph leaves it. Every inactive component is already a connected component of the final graph.
Since at most \(K\) edges of the historical spanning forest are missing at the end of the stream, the surviving forest has at most \(2K\) active components. From this point, \(O(\log K)\) contraction rounds suffice for constant error (\cref{app:rand-upper-proof}).

Second, we reduce the space needed to find an edge leaving a component. In each round, AGM samples edges at \(O(\log n)\) geometrically decreasing rates. We retain only \(O(\log K)\) sampling levels.
Fix an active component \(C\), and let \(s\) be the number of edges leaving \(C\) in the final graph. If \(s\le K\), one of the retained levels samples exactly one of these edges with constant probability~\cite{jowhari_et_al:PODS.2011}.
If \(s>K\), then among the historical edges leaving \(C\), \(s\) are present and at most \(K\) are missing. Under approximately min-wise independent ranks~\cite{saks_et_al:IPL.2000}, the lowest-ranked historical edge leaving \(C\) is therefore one of the \(s\) present edges with constant probability. By the cut property, the historical minimum spanning forest for these ranks contains this edge.

The basic algorithm gives the constant-error bound in \cref{thm:main}. For any fixed \(c>0\), our second algorithm attains the optimal space bound in that theorem for every target failure probability \(n^{-c}\le\delta\le1/3\). Set \(B=\lceil\log((K+1)/\delta)\rceil\).
When \(K\le B\), we maintain \(K+1\) forests. On insertion, we place an unstored edge in the first forest where its endpoints are disconnected, if there is such a forest.
The surviving edges in these forests preserve the connectivity of the final graph, giving a deterministic algorithm using \(O(n(K+1)\log n)\) bits~\cite{nagamochi1992linear}.
When \(K>B\), all \(O(B)\) contraction rounds share one Vandermonde sketch per vertex for cuts of size at most \(B\)~\cite[Appendix~E.1]{feng_et_al:ICML.2024}.
Each round then uses \(O(\log(1+K/B))\) sampling levels with fresh hashes for final cuts of size \(B<s\le K\), and independently ranked historical minimum spanning forests for larger cuts.
The algorithm achieves failure probability at most \(\delta\) using \(O_c\!\left(nB\log n\,\log\!\left(1+\frac{K+1}{B}\right)\right)\) bits (\cref{thm:high-prob-upper}).
For \(1\le k\le n^2\), this gives the upper bound in the second part of \cref{thm:main}.

\paragraph*{Lower Bound for Randomized Algorithms.} 
To establish the randomized space lower bound in the bounded-deletion, our proof adapts the framework of Nelson and Yu \cite{nelson_yu:SODA.2019} in proving the optimal space lower bound for spanning forest computations in the turnstile models. The adaptation of their lower bound argument to the bounded-deletion setting involves an additional stage beyond their proof paradigm and refined parameterizations to accommodate the bounded-deletion constraints.

We first outline the three-stage proof of \cite[Theorem 4.1]{nelson_yu:SODA.2019} for the optimal turnstile streaming lower bound for spanning forests. The starting point of the lower bound argument stems from the communication problem of \emph{subset universal relation}. In this promise problem, Alice and Bob receive two sets $A$ and $B$ from the universe $[U]$ under the promise that $B\subsetneq A$, and Bob's goal is to output an element $x\in A\setminus B$ upon receiving a single message from Alice.
In the first stage of their proof, \cite[Lemma 4.1]{nelson_yu:SODA.2019} proves the reduction of the spanning forest problem from the communication problem of $n$-fold universal subset relation $(\UR)^{\otimes n}$, which the players solve $n$ independent instances of $\UR$ (see \cref{eq:nfold} for the formal definition).
Second, using a direct-product-type argument based on \cite{Braverman_et_al:FOCS.2013}, \cite[Lemma 4.3]{nelson_yu:SODA.2019} converts a one-way randomized protocol for $(\UR)^{\otimes n}$ with cost $C$ and error $\delta$ into a \emph{lossy protocol} (to be formally defined in \cref{def:lossy}) for $\UR$ with cost $C/n$ and error $O(\sqrt{\delta/n})$. Finally, adapting an encoding argument \cite[Theorem 2]{kapralov_et_al:FOCS.2017}, \cite[Lemma 4.5]{nelson_yu:SODA.2019} establishes an internal information cost lower bound for such lossy protocols for $\UR$.

To adapt the framework of Nelson and Yu to the bounded-deletion setting, a quick inspection reveals that by emulating the reduction, the $k$-deletion spanning forest problem admits a reduction from $(\UR)^{\otimes n}$ with the additional constraint that Bob's input $(B_1,\ldots,B_n)$ satisfies $\sum_{i=1}^n \abs{B_i} \leq k$.
Directly incorporating this constraint would disrupt the product structure of $(\UR)^{\otimes n}$, thus requiring substantial modifications to the arguments. 
The most crucial modification to their proof framework is the additional $\INDEX$-lifting stage. The $\INDEX$ gadget lifting technique has been employed in prior work (e.g., \cite{khanna_konrad:ITCS.2022, khanna_konrad_dark:ICALP.2025}) to embed a hard communication problem among multiple independent instances. This embedding retains a large multi-fold size-restricted $\UR$ instance (see \cref{def:size-restricted-UR} for the precise definition) for a strong lower bound, and bypasses the complications of individually restricting Bob's input size in a multi-fold $\UR$ problem.

By adopting a structure of an $\INDEX_g$-lifting of $(\UR)^{\otimes m}$, the deletion bound imposes a trade-off constraint on $g$ and $m$ in the reduction stage. 
This constraint prevents us from directly applying \cite{nelson_yu:SODA.2019} with their original parameter choices.
To attain the best possible lower bound from the information complexity analysis under a prescribed deletion bound, we carefully track the dependence on the relevant parameters. This refinement also yields improvements to the original bounds over a broader range of parameter regimes, independently of our bounded-deletion application, as exemplified by \cref{lem:direct-product}. In particular, we improve the error dependence of the direct-product lemma for lossy protocols from a square-root dependency to a linear dependence, potentially making the lemma useful in a broader range of applications. Finally, by a careful inspection of the proof in \cite{nelson_yu:SODA.2019} for retrieving the relevant parameters, we obtain the complete general space lower bound in \cref{thm:general_LB} for varying deletion bound and error bound.

\subsection{Applications}

Our results in the bounded-deletion model can be applied naturally to turnstile streaming algorithms that use spanning-tree computation as a primitive, with careful control of the failure probabilities across subroutine invocations. Such applications must also account for the deletion bound of each derived stream.

Let $S(n)$ denote the space required for single-pass spanning-tree computation, returning an edge-maximal spanning forest on disconnected inputs. Computing a $\lambda$-edge-connectivity certificate uses $\lambda+1$ independent spanning-forest sketches and $O(\lambda S(n))$ space~\cite{ahn_et_al:SODA.2012}. Computing a $\kappa$-vertex-connectivity certificate uses $O(\kappa^2\log n)$ spanning-forest instances on sampled vertex subsets and at most $O(\kappa^2\log n\cdot S(n/\kappa))$ space~\cite{Assadi23}. Bipartiteness testing reduces to computing one edge-maximal spanning forest on the bipartite double cover, using $O(S(2n))$ space~\cite{ahn_et_al:SODA.2012}.

Spanning-tree computation also supports algorithms for specialized spanning trees. A sparsifier preserving at least a $(1-\varepsilon)$ fraction of the maximum number of leaves in a spanning tree can be constructed using one spanning-tree computation and additional incident-edge sampling, using $O(S(n))+\tilde{O}_\varepsilon(n)$ space~\cite{ChangFHT20}. For positive weights in $[1,W]$ represented using $O(\log n)$ bits, computing a $(1+\varepsilon)$-approximate minimum spanning tree reduces to $O(1+\varepsilon^{-1}\log W)$ parallel spanning-tree instances, requiring $O((1+\varepsilon^{-1}\log W)S(n))$ space~\cite{CrouchMS13}. All these constructions use a single pass.


\section*{Acknowledgment}

This manuscript is based on our October submission to STACS 2027. During the preparation of this arXiv version, independent and concurrent work on spanning forests in the bounded-deletion model was posted on arXiv \cite{CCMW26}. Our bounds, obtained using different techniques, match theirs for error probabilities down to inverse-polynomial levels.

This work is written with assistance from large language models through paid plans between July and
November 2026. Claude Opus 5.5 (Anthropic) assisted in drafting
\Cref{sec:rand-upper,app:rand-upper-proof} and parts of the introduction, and in completing
the analysis of \Cref{app:dynamic} and editing it for readability.
GPT-6-Astra (OpenAI) assisted in refining the algorithm for target failure probabilities
\(n^{-c}\le\delta\le 1/3\) in the second part of \Cref{thm:main} and in drafting
\Cref{sec:high-prob-upper}. It is also used in proof checking and literature review in \Cref{sec:kec-deter}.
GPT-6.1 Sol (OpenAI) assisted in the presentation of the analysis of
\Cref{sec:LB,sec:direct-product,app:dynamic,app:Lemma5}. 
The authors reviewed all AI-assisted content,
verified every algorithm, proof, and cited result against the original
sources, and take full responsibility for the content of this paper.

\section{Preliminaries}
\label{sec:preliminaries}

\paragraph*{Notations.} For an integer $N$, we write $[N]=\set{1,\ldots,N}$. For any set $S$ and $r\leq \abs{S}$, we denote $\binom{S}{r}$ and $\binom{S}{\leq r}$ the collections of subsets of $S$ of size exactly $r$ and at most $r$ respectively.

All unqualified logarithms are to base two, and $\ln$ denotes the natural logarithm. In bounds stated up to constant factors, we may abbreviate $\log_2(1+x)$ by $\log x$ when $x\ge 1$ throughout the stated parameter range. An explicitly written $\log(1+x)$ always denotes the ordinary base-two logarithm. In exact definitions, logarithms have their usual meaning.

For $p\in (0,1)$, we define $\L(p)\defeq \log(1/p)$, and denote the binary entropy function by $\ent(p)\defeq p\L(p)+(1-p)\L(1-p)$.

\paragraph*{Graph Streams.}
Let \(V=[n]\) be a fixed vertex set and let
\(
  \mathcal{E}:=\binom{V}{2}, 
  N:=|\mathcal{E}|=\binom{n}{2}.
\)
A graph stream is a sequence of updates
\(
  (e,\Delta)\in\mathcal{E}\times\{-1,+1\},
\)
where \(\Delta=+1\) denotes an insertion and \(\Delta=-1\) denotes a deletion. We consider valid simple-graph streams: an edge can be inserted only when it is absent and deleted only when it is present. An edge may be deleted and subsequently reinserted.

Let \(G=(V,E)\) denote the final graph of the stream. The algorithm is given an upper bound \(k\) on the total number of deletion updates. Throughout the paper, we use the capped parameter
\(
  K:=\min\{k,N\}.
\)

The update sequence is \emph{oblivious}: it is fixed independently of the private randomness of the streaming algorithm. A randomized one-pass spanning forest algorithm has an error of at most \(\delta\) if, for every valid oblivious stream, it outputs a spanning forest of the final graph $G$ with probability at least \(1-\delta\), where the probability is over its private randomness.
All space bounds are measured in bits and include stored random seeds and the working memory used during postprocessing and output construction. 


\paragraph*{Communication Complexity.}
For our lower-bound results, we appeal to communication complexity as is customary in the streaming literature. Since we focus on one-pass streaming lower bounds, we consider exclusively one-way two-party communication. 
By convention, the two parties are Alice and Bob, where Alice sends a single message to Bob, who then produces the output. For a communication problem $F$, the deterministic one-way communication complexity and the (public-coin) randomized one-way communication complexity with error at most $\delta$ are denoted by $\Dcc(F)$ and $\Rcc_\delta(F)$ respectively.

Given a communication problem $F:\calX\times\calY\to\calZ$, we introduce two operations for constructing new communication problems from $F$. The first is the \emph{$n$-fold product} of $F$, denoted by $F^{\otimes n}:\calX^n\times \calY^n\to \calZ^n$, defined as 
\begin{equation}\label{eq:nfold}
    F^{\otimes n}((x_1,\ldots,x_n),(y_1,\ldots,y_n)) \defeq (F(x_1,y_1),\ldots,F(x_n,y_n)).
\end{equation}
The second operation is \emph{$\ell$-$\INDEX$ lifting}, which defines the communication problem $\INDEX_\ell\circ F:\calX^\ell\times (\calY\times [\ell])\to \calZ$ by
\begin{equation}\label{eq:index-lift}
    \INDEX_\ell\circ F((x_1,\ldots,x_\ell), (y, J)) \defeq F(x_J,y).
\end{equation}

After establishing reductions from communication problems to streaming problems, our lower-bound arguments proceed with information complexity serving as a central technical tool in the subsequent analysis.
The notations and preliminaries on information complexity are deferred to \cref{sec:IC}.
\section{Randomized Upper Bound}
\label{sec:rand-upper}


For a graph \(J=(V,E_J)\) and a set \(W\subseteq V\), let \(\delta_J(W)\coloneqq\{uv\in E_J:|\{u,v\}\cap W|=1\}\) denote the set of edges of \(J\) with exactly one endpoint in \(W\).

We prove the upper bound in the first part of \cref{thm:main}, with the following more precise error guarantee.

\begin{theorem}
\label{thm:rand-upper}
Let \(c>0\) be a constant.
There is a randomized one-pass streaming algorithm that, for every \(n\ge2\) and every deletion bound \(k\ge1\), outputs a spanning forest of the final graph \(G\) with error at most \((K+1)^{-c}+n^{-c}\) and uses \(O\left(n\log n\log^2K\right)\) bits of space.
\end{theorem}

For any fixed \(0<\delta\le1/3\), choosing \(c\) sufficiently large makes the failure probability at most \(\delta\), establishing the claimed upper bound.

For \(n\le1\), output the empty forest.
If \(k=0\), the stream is insertion-only, so retaining each edge that joins distinct components of the maintained forest gives a deterministic one-pass algorithm using \(O(n\log n)\) bits.
Henceforth, fix \(c>0\) and assume \(n\ge2\) and \(k\ge1\). Constants hidden in \(O(\cdot)\) may depend only on \(c\).

\subsection{Overview}
\label{subsec:rand-upper-overview}

We use the contraction procedure of Ahn, Guha, and McGregor (AGM)~\cite{ahn_et_al:SODA.2012}, with a different initial forest and different structures for recovering outgoing edges.

\begin{center}
\fbox{\begin{minipage}{0.94\linewidth}
\textbf{Contraction procedure}\par\smallskip
\textit{Input:} a forest \(\mathcal F\subseteq E\), and recovery structures \(1,\ldots,R\) maintained during the stream.
\begin{enumerate}
    \item \textbf{for} \(r=1,\ldots,R\) \textbf{do}
    \begin{enumerate}
        \item Query structure \(r\) with each component of \(\mathcal F\) for an edge of \(G\) that leaves the component, and collect the returned edges as \emph{candidates}.
        \item Scan the candidates, adding each edge to \(\mathcal F\) if its endpoints lie in distinct components of the current \(\mathcal F\).
    \end{enumerate}
    \item \textbf{return} \(\mathcal F\).
\end{enumerate}
\end{minipage}}
\end{center}

A failed query contributes no candidate.

The AGM sketch assigns each vertex \(v\) a \emph{signed incidence vector} \(x^{(v)}\). Each edge of \(G\) has opposite signs at its two endpoints. In \(\sum_{v\in C}x^{(v)}\), the two contributions of each internal edge cancel, while each edge of \(\delta_G(C)\) contributes exactly once (\cref{eq:rand-upper-cut-cancellation}). By linearity, summing the vertex sketches lets an \(\ell_0\)-sampler recover an outgoing edge with constant probability whenever the cut is nonempty. Starting from the empty forest, AGM uses \(O(\log n)\) rounds. At \(O(\log^2 n)\) bits per vertex and round~\cite{jowhari_et_al:PODS.2011}, this costs \(O(n\log^3 n)\) bits.

Let \(H\) be the historical graph of all edges ever inserted. It only grows, and at most \(K\) of its edges are absent from \(G\) (\cref{subsec:rand-upper-backbone}). We use both facts to modify AGM as follows.

\begin{enumerate}
    \item \textit{Fewer rounds.}
    We maintain a spanning forest \(F\) of \(H\). Each edge of \(F\) has its own \emph{presence bit}, which is one if that edge is present in the current graph and zero otherwise.
    At the end of the stream, these bits identify the surviving edges \(F_0\coloneqq F\cap E\).
    By \cref{lem:rand-upper-backbone}, at most \(2K\) components of \(F_0\) are active. Starting from \(F_0\), the contraction needs only \(O(\log K)\) rounds (\cref{subsec:rand-upper-contraction}).
    \item \textit{Truncated sampling for small cuts.}
    An \(\ell_0\)-sampler subsamples the coordinates at rates \(2^{-1},2^{-2},\ldots\) down to \(2^{-\Theta(\log N)}\) and looks for a rate at which exactly one coordinate of the support survives.
    We keep only \(L=O(\log K)\) rates, which is enough for every component whose final cut \(\delta_G(C)\) has at most \(K\) edges.
    At each rate, we store only two linear measurements over a prime field. Using a randomized fingerprint, we test whether the subsampled vector has exactly one nonzero coordinate and return a candidate edge on acceptance.
    A vertex thus stores \(O(\log n\log K)\) bits per round instead of \(O(\log^2n)\).
    \item \textit{Historical minimum spanning forests for large cuts.}
    A component whose final cut has more than \(K\) edges is not covered by the truncated sampler.
    For such a component, however, at most \(K\) edges of the historical cut \(\delta_H(C)\) are missing from the final graph, so more than half of the historical cut survives.
    We draw a permutation from an approximately min-wise independent family~\cite{saks_et_al:IPL.2000} and maintain the minimum spanning forest of \(H\) under the resulting edge ranks.
    By the cut property, the historical minimum spanning forest contains the lowest-ranked edge of \(\delta_H(C)\). With constant probability, this edge belongs to \(\delta_G(C)\).
    A constant number of independent such forests per round replaces the sampler for large cuts.
\end{enumerate}

Altogether, a round costs \(O(n\log n\log K)\) bits, and \(O(\log K)\) rounds give the space bound of \cref{thm:rand-upper}.
When \(K=N\), the bound is \(O(n\log^3n)\), the same as for AGM sketch.

To query a component, the algorithm applies the fingerprint test at the retained sampling levels and returns the first candidate it finds. If all tests reject, it scans the historical forests for a present edge leaving the component.
Since the algorithm does not know which components are active or whether their cuts are small or large, it applies the same query to every component.

The recovery structures and full proof of \cref{thm:rand-upper} appear in \cref{app:rand-upper-proof}. First suppose the fingerprint tests are exact. In each round, fresh sampling hashes and edge ranks give each active component a constant probability of finding an outgoing edge. The expected number of active components thus drops by a constant factor per round, so the probability that at least one active component remains after \(R\) rounds is at most \((K+1)^{-c}\) (\cref{lem:rand-upper-convergence}). Fingerprint errors add at most \(n^{-c}\) to the failure probability.

\subsection{The historical backbone}
\label{subsec:rand-upper-backbone}

Denote \(H=(V,E_H)\) the historical graph.
Then \(E\subseteq E_H\).
Let \(Z\coloneqq E_H\setminus E\).
Every edge in \(Z\) was deleted at least once, so \(|Z|\le k\).
Since \(|Z|\le N\) as well, we have 
\begin{equation}
    |Z|\le K.
    \label{eq:rand-upper-missing}
\end{equation}

We maintain a spanning forest \(F\) of the historical graph by adding each inserted edge whose endpoints lie in distinct components of \(F\). Edges never leave \(F\).
Each edge of \(F\) has a presence bit, set to one on insertion and zero on deletion.
At the end of the stream, these bits identify \(F_0\coloneqq F\cap E=F\setminus Z\), a forest of \(G\).
The contraction starts from \(F_0\), which is missing at most \(K\) edges of \(F\).

In cut notation, a component \(C\) of a forest \(T\subseteq E\) is active exactly when \(\delta_G(C)\neq\varnothing\).

The next lemma bounds the number of active components of the backbone and shows that the contraction may stop once no component is active.

\begin{lemma}
\label{lem:rand-upper-backbone}
The forest \(F_0\) has at most \(2K\) active components.
Moreover, any forest \(T\subseteq E\) with no active component is a spanning forest of \(G\).
\end{lemma}

\begin{proof}
Let \(C\) be an active component of \(F_0\), and let \(uv\) be a final edge leaving \(C\), with \(u\in C\).
Since \(uv\in H\), the historical spanning forest \(F\) contains a path from \(u\) to \(v\).
The first edge of this path leaving \(C\) belongs to \(F\setminus F_0\).
Thus every active component contains an endpoint of an edge in \(F\setminus F_0\), so there are at most \(2|F\setminus F_0|\le 2K\) active components.

If no edge of \(G\) leaves any component of a forest \(T\subseteq E\), then its components are precisely the connected components of \(G\), so \(T\) is a spanning forest of \(G\).
\end{proof}

\section{Randomized Lower Bound}\label{sec:LB}
In this section, we establish a matching one-pass randomized lower bound for spanning forest algorithms in the bounded-deletion streaming model. 


In the following subsections, we describe the modifications needed to establish our bounded-deletion lower bound. Our analysis relies primarily on tools from information complexity, which we present the relevant preliminaries in \cref{sec:IC}.

\subsection{Reduction from $\INDEX$ Embedding of Multi-fold $\UR$}
Our lower bound argument begins with a reduction from an $\INDEX$-lift of a size-restricted subset universal relation problem.
\begin{definition}\label{def:size-restricted-UR}
    Let $U\in \N$ and $1\leq \alpha \leq U$. The \emph{size-restricted subset universal relation} $\UR_{U;\alpha}$ refers to the following one-way two-party communication problem. Alice receives $A\in \binom{[U]}{\alpha}$ and Bob receives $B\in \binom{A}{\leq \alpha-1}$, after receiving a single message from Alice, Bob must output an element in $A\setminus B$.
\end{definition}

We set up some notation and definitions needed for our reduction from streaming algorithms to information complexity.
For a one-way protocol $\Pi$, we use $M_\Pi$ and $R_\Pi$ to denote its message and public randomness respectively, without explicit introduction. We omit the subscripts whenever the protocol is clear from context.

Our reduction lemma yields a bound on the internal information cost of a one-way protocol, which is defined as follows:
\begin{definition}[Internal Information Cost]
    Let $\Pi$ be a one-way protocol, $(A,B)\sim \mu$ be a joint input distribution. The \emph{internal information cost} of $\Pi$ under $\mu$ is defined as
    \(
        \I_\mu(A;M_\Pi\mid BR_\Pi).
    \)
\end{definition}
This definition coincides with the standard notion of internal information cost for interactive communication protocols \cite{BBCR10,BYJKS02}, specialized to one-way protocols.

\begin{lemma}\label{lem:reduction}
    Suppose there exists a randomized $k$-deletion streaming algorithm for spanning forests on $2n$-vertex graphs using $C$ bits of space with error probability at most $\delta$. Then, for every $g\in[n/2]$, letting $m=\floor{n/g}$ and $\alpha\in \N$ such that $m(\alpha-1)\leq K$, there exists a one-way protocol for $(\UR_{n;\alpha})^{\otimes m}$ with internal information cost at most $C/g$ and error probability at most $\delta$.
\end{lemma}
\begin{proof}
    We form a bipartite graph $G$ with $n$ vertices on each side as follows. On the left-hand side, we form $g$ disjoint groups of size $m$ (and leave the remaining vertices isolated), labelling the vertices as $\ell^i_{j}$ for $i\in[g]$ and $j\in[m]$. 
    For the right-hand side, we label the vertices as $r_j$ for $j\in [n]$.    
    Writing $\Phi=(\UR_{n;\alpha})^{\otimes m}$, we first show that the spanning forest streaming problem is reduced to $\INDEX_g\circ \Phi$.

    Let Alice's input be $(A_j^i)_{i\in [g],j\in [m]}$, where $\vec{A}^i=(A_1^i,\ldots,A_m^i)\in \binom{[n]}{\alpha}^m$ forms Alice's input to the $i$-th copy of $\Phi$ in $\INDEX_g\circ \Phi$. For each $i\in[g],j\in [m]$, Alice inserts the edge between $\ell^i_{j}$ and $r_a$ in the graph stream for each $a\in A_j^i$.\footnote{To enforce the additional connectedness promise, Alice can additionally insert a fixed spanning tree with root $r_1$ and children $r_j$ for $j\in [2,n]$ and the $n-gm$ unused left vertices. These additional edges ensure that the final graph is connected since every vertex $\ell^i_j$ retains at least one incident edge after Bob's deletion.} After all the insertions, she sends the memory state of the simulated streaming algorithm to Bob. On his input $I\in [g]$ and $\vec{B}=(B_1,\ldots,B_m)\in \prod_{j=1}^m\binom{A^I_j}{\leq \alpha-1}$, Bob deletes the edge between $\ell^I_{j}$ and $r_{b}$ for each $j\in [m]$ and $b\in B_j$. The number of edges deleted is at most $m(\alpha-1)\leq K$.
    After the deletions, the algorithm returns a spanning forest $F$ of $G$. Since $B_j\subsetneq A^I_j$ for each $j\in [m]$, for a successful run of the streaming algorithm, Bob can retrieve $x_j\in A^I_j\setminus B_j$ for each $j$ by finding a surviving edge connected to $\ell^I_{j}$, thus $(x_1,\ldots,x_m)$ forms a valid output to $\Phi(\vec{A}^I,\vec{B}) = \INDEX_g\circ \Phi((\vec{A}^1,\ldots,\vec{A}^g), (I,\vec{B}))$. This reduction also yields a communication protocol $\Pi$ for $\INDEX_g\circ \Phi$ with cost $C$ and error probability at most $\delta$.

    Next, we show that a one-way protocol $\Pi$ for
    $\INDEX_g\circ\Phi$ with communication cost $C$ and
    worst-case error at most $\delta$ implies a one-way
    protocol $\Pi'$ for $\Phi$ with the same error guarantee
    and internal information cost at most $C/g$ under any
    input distribution. This follows from a now-standard
    information-theoretic argument for $\INDEX$.

    Let $\mu$ be any distribution on inputs
    $(\vec{A},\vec{B})$ to $\Phi$, and let $\mu^A$ denote
    its marginal on Alice's input. Independently sample
    $\vec{A}^1,\ldots,\vec{A}^g$ from $\mu^A$. By the chain rule and independence of the $g$ blocks,
    \[
        C\geq \I(\vec{A}^1\ldots \vec{A}^g;M_{\Pi} \mid R_{\Pi}) =  \sum_{i=1}^g \I(\vec{A}^i; M_{\Pi} \mid \vec{A}^{<i}R_{\Pi}) \geq \sum_{i=1}^g \I(\vec{A}^i;M_{\Pi} \mid R_{\Pi}).
    \]
    Hence there exists $i^*\in [g]$ such that $\I(\vec{A}^{i^*};M_{\Pi} \mid R_{\Pi})\leq C/g$. 
    Now we construct $\Pi'$ as follows. On input
    $(\vec{A},\vec{B})$ to $\Phi$, Alice forms
    $\widetilde{A}$ by placing $\vec{A}$ in the
    $i^*$-th block and independently sampling the
    remaining blocks from $\mu^A$ using private
    randomness. The two parties then run $\Pi$
    on inputs $\widetilde{A}$ and $(i^*,\vec{B})$,
    respectively.

    It is clear that the worst-case error guarantee of $\Pi'$ is the same as that of $\Pi$. Moreover, 
    \[
        \I_{\mu}(\vec{A};M_{\Pi'}\mid \vec{B} R_{\Pi'})\leq \I_{\mu}(\vec{A};M_{\Pi'}\mid R_{\Pi'}) = \I(\vec{A}^{i^*};M_{\Pi}\mid R_{\Pi}) \leq \frac{C}{g}.
    \]  
    Here the first inequality follows from \cref{lem:one-way}. This establishes that for every input distribution \(\mu\) for \(\Phi\), there exists a one-way protocol \(\Pi'\) for \(\Phi\) with the desired worst-case error and information cost bound.
\end{proof}

\subsection{Direct-Product Lemma for Lossy Protocols for $\UR$}
The next step in our analysis is to adapt \cite[Lemma 4.3]{nelson_yu:SODA.2019} to our setting. The key modification is to replace the assumption on the protocol's communication cost with an internal information complexity bound. We also reorganize the original proof to explicitly track the parameters relevant to our setting, while streamlining and simplifying the presentation of the argument.

\begin{definition}[Lossy Protocol]\label{def:lossy}
    Let $\mu$ be an input distribution for a two-player communication problem. A \emph{lossy protocol} $\tilde\Pi$ consists of a randomized message generator and a randomized decoder. On input $(x,y)$, the message generator samples a message $m$ from a conditional distribution $\Msg(\cdot\mid x,y)$. Given $(m,y)$, the decoder samples an output $z$ from a conditional distribution $\Dec(\cdot\mid m,y)$. The resulting joint distribution is given by
    \[
        \rho(x,y,m,z) = \mu(x,y)\Msg(m\mid x,y)\Dec(z\mid m,y).
    \]
    The \emph{forward information cost} of $\tilde\Pi$ under $\mu$ is defined as $\I_\rho(X;M\mid Y)$, and the \emph{backward information cost} of $\tilde\Pi$ under $\mu$ is defined as $\I_\rho(Y;M\mid X)$.

    A \emph{$\theta$-lossy protocol} is a lossy protocol with backward information cost at most $\theta$.
\end{definition}
As remarked in \cite{nelson_yu:SODA.2019}, a 0-lossy protocol is a standard one-way communication protocol.

\begin{lemma}[Direct-Product Lemma for Lossy Protocol]\label{lem:direct-product}
    Let $m\geq 2$, $\mu$ be an input distribution for $f:\calX\times \calY\to \calZ$. Suppose $\Pi$ is a one-way protocol computing $f^{\otimes m}$ under $\mu^{\otimes m}$, such that the success probability is at least $p\in (0,1)$, and
    \[
        \I(\vec{X};M_\Pi\mid \vec{Y}R_\Pi)\leq \gamma.
    \]
    where $(\vec{X},\vec{Y})\sim \mu^{\otimes m}$.    
    Then there exist an input distribution $\mu'$ and an $O(\L(p)/m)$-lossy protocol $\tilde{\Pi}$ for $f$ under $\mu'$ with joint distribution $\rho$ that satisfy the following properties.
    \begin{enumerate}
        \item \emph{Protocol guarantees}: The forward information cost and error probability satisfy
        \[
            \I_\rho(X;M\mid Y)=O\left(\frac{\gamma+\ent(p)}{mp}\right) 
            \quad\text{and}\quad
            \Pr_{\rho}[f(X,Y)\neq Z] = O\left(\frac{\L(p)}{m}\right).
        \]
        \item \emph{Input distribution guarantees}: $\mu'$ is $(O(\L(p)),O(\L(p)/m))$-KL-close to $\mu$, where the closeness notions are defined as follows:
        \begin{itemize}
            \item $\nu'$ is $\zeta$-close to $\nu$ if $\KL(\nu'\Vert \nu)\leq \zeta$;
            \item $\nu'$ is $\eta$-conditionally-close to $\nu$ on $(X,Y)$ if 
            \[
                \CRE{\nu'[X\mid Y]}{\nu[X\mid Y]}{\nu'[Y]}\leq \eta
                \quad\text{and}\quad
                \CRE{\nu'[Y\mid X]}{\nu[Y\mid X]}{\nu'[X]}\leq \eta.
            \]
            \item $\nu'$ is $(\zeta,\eta)$-KL-close to $\nu$ if $\nu'$ is $\zeta$-close to $\nu$ and $\eta$-conditionally-close to $\nu$ on the input variables.
        \end{itemize}
    \end{enumerate}
\end{lemma}
The proof closely follows the approach of \cite{nelson_yu:SODA.2019} with two main modifications. First, we replace the original communication-cost bound for the multi-fold protocol with an internal information-cost bound $\gamma$. This requires a modified analysis of the forward information cost of the resulting lossy protocol using a different mutual information argument. The modified bound \cref{eq:FC_bound} is proven in \cref{subsec:config-sampling}.

Second, we improve the error bound of the lossy protocol from $O(\sqrt{\L(p)/m})$ to $O(\L(p)/m)$ by applying the data-processing inequality for KL divergence to the correctness event instead of Pinsker's inequality. The improved bound \cref{eq:err-bound} is proven in \cref{subsec:input-distr}.

Our reformulation also aims to make the argument more natural by aligning the quantities being bounded more closely with the information-cost and correctness guarantees required of the lossy protocol, thus providing a clearer motivation for this otherwise somewhat artificial notion.
We defer the complete proof to \cref{sec:direct-product}.

\subsection{Lossy Protocol Lower Bound for $\UR_{U;\alpha}$}
With a direct-product lemma for lossy protocols in place, the remaining task is to derive a lossy protocol cost lower bound for $\UR_{U;\alpha}$. For our use case for bounded-deletion streaming lower bound with varying error bound and deletion bound, we require maximal flexibility of parameter choices.

In \cite[Lemma 4.5]{nelson_yu:SODA.2019}, the hard distribution $\mu^{\UR}$ implicitly enforces that Alice's and Bob's sets are of size at most $\alpha = \sqrt{U\L(\delta)}$.\footnote{Some parameter definitions in our presentation differ from those in \cite{nelson_yu:SODA.2019} despite the same symbols. Throughout the subsequent discussion, we adhere to the notation introduced in this paper.} 
For a $\theta$-lossy protocol for $\UR_{U;\alpha}$ with error probability at most $\delta'$ on an input distribution $\tilde{\mu}$ $(O(1),\eta)$-KL-close to $\mu^{\UR}$ (see \cref{lem:direct-product} for the definition of KL closeness), a
direct adaptation of the encoder-decoder argument of \cite[Lemma 4.5]{nelson_yu:SODA.2019} yields a forward information cost lower bound of 
\[
    \Omega\left(\L(\delta') \log\frac{\alpha}{\L(\delta')} \log\frac{U}{\alpha} \right) 
\]
as long as $\eta=O((\delta')^2),\alpha\in [\Omega(\L(\delta')^3), O(U)],\delta'\in [2^{-U^{1-O(1)}},O(1)]$. By setting $U=n$ and applying \cref{lem:reduction,lem:direct-product} combined, this shows that for a spanning forest streaming algorithm with error bound $\delta$ and deletion bound $K$, its space usage is at least 
\[
    m\cdot g\cdot \Omega\left(\L\left(\frac{\delta}{m}\right) \log\frac{\alpha}{\L(\delta/m)} \log\frac{n}{\alpha} \right) = \Omega\left( n\log \frac{m}{\delta} \log\frac{\alpha}{\L(\delta/m)} \log\frac{n}{\alpha} \right)  
\]
with the additional restriction that $m\alpha=O(K)$. In particular, assuming all parameters are within the valid range, for $\delta=O(1)$, choosing $\alpha=m=\Theta(\sqrt{K})$ yields a lower bound of $\Omega(n\log n \log^2 K)$.

Certainly, the apparent issue is that the above choice of $\alpha$ and $m$ may yield parameters outside the permitted range when adopting their proof. We handle this issue with a fully parameterized information lower bound for lossy protocols for $\UR_{U;\alpha}$.
\begin{restatable}{lemma}{LBUR} \label{lem:LB-UR}
    There exist absolute constants $c,h_0>0$ such that for every $U,\alpha\in \N$ and $h\in \R$ satisfying $h_0\leq h\leq \alpha\leq U/2$, there is an input distribution $\mu=\mu_{U,\alpha,h}$ for $\UR_{U;\alpha}$ with the following properties.
    Let $\mu'$ be $(\zeta,\eta)$-KL-close to $\mu$, and $\tilde{\Pi}$ be a $\theta$-lossy protocol under $\mu'$ with error probability at most $\epsilon$. If $\epsilon+\eta+\theta\leq 2^{-h}$, then the forward information cost of $\tilde{\Pi}$ is at least $ch\log(\alpha/h) \log(U/h)$.
\end{restatable}
The proof of \cref{lem:LB-UR} follows the same proof in \cite{nelson_yu:SODA.2019} with careful parameter tracking and is deferred to \cref{app:Lemma5}.
Finally, by combining \cref{lem:reduction,lem:direct-product,lem:LB-UR}, we establish our streaming space lower bound generalized to varying deletion bound and error bound. 
\begin{theorem}\label{thm:general_LB}
    Let $k\geq 1$ and $0<\delta\leq 1/3$. Define 
    \[
        K'=\min\set{k,n} 
        \quad\text{and}\quad 
        \Gamma = \min\setB{K',\L(\delta)+\log K'}.
    \]
    Every randomized one-pass spanning forest streaming algorithm on $2n$-vertex graphs with deletion bound $k$ and error probability at most $\delta$ uses
    \[
        C = \Omega\left(n\Gamma \log\frac{K'}{\Gamma} \log\frac{n}{\Gamma}\right) 
    \]
    bits of space.
\end{theorem}
\begin{proof}
    Let $C''$ be a sufficiently large universal constant. If $\log K'\leq C''$, the claimed bound reduces to $C=\Omega(n\log n)$, which follows immediately from the insertion-only lower bound.

    Now assume $\log K'>C''$. For $g\in [n/2]$, let $m=\floor{n/g}$. Write $p=1-\delta$. Let $\alpha\in \N$ and $h\in\R$ satisfy
    \begin{equation}\label{eq:parameters}
        m(\alpha-1)\leq K,\qquad h_0\leq h\leq \alpha\leq n/2,\qquad \frac{C_0\L(p)}{m}\leq 2^{-h}
    \end{equation}
    for some sufficiently large constant $C_0>0$. 
    
    Let $\mu=\mu_{n,\alpha,h}$ be the distribution given in \cref{lem:LB-UR}. Applying \cref{lem:reduction,lem:direct-product} yields an $O(\L(p)/m)$-lossy protocol $\tilde{\Pi}$ for $\UR_{n;\alpha}$ and an input distribution $\mu'$, such that $\tilde{\Pi}$ has an error bound $O(\L(p)/m)$ and forward information cost 
    \[
        O\left(\frac{C/g+\ent(p)}{mp}\right) = O\left(\frac{C}{n}+\frac{1}{m}\right)
    \]
    under $\mu'$, and $\mu'$ is $(O(\L(p)), O(\L(p)/m))$-KL close to $\mu$. By \cref{lem:LB-UR}, we have\footnote{To avoid cluttering big-O notation, here we use the notation $f\gtrsim g$ to denote $f=\Omega(g)$.}
    \[
        \frac{C}{n}+\frac{1}{m} \gtrsim h\log\frac{\alpha}{h} \log\frac{n}{h} 
        \implies C  \gtrsim n h\log\frac{\alpha}{h} \log\frac{n}{h}  - \frac{n}{m}.
    \]
    We consider two cases:
    \begin{itemize}
        \item \emph{$\L(\delta)\geq \log K'$}: 
        Let $h' = \min\set{K',\L(\delta)}$. Then $\Gamma\in [h',2h']$. Choose 
        \[
            g= \floor{n/2}, \qquad m=2, \qquad \alpha = \floor{K'/8}+1, \qquad h=c'h',
        \]
        where $c'>0$ is a sufficiently small constant such that \cref{eq:parameters} is satisfied. The declared lower bound for $C$ follows from the facts that $\alpha=\Theta(K')$ and $h=\Theta({\Gamma})$.
        \item \emph{$\L(\delta)< \log K'$}: 
        Choose
        \[
            g= \floor{n/\sqrt{K'}}, \qquad m=\floor{n/g}=\floor{\sqrt{K'}}\pm O(1), \qquad \alpha = \floor{\sqrt{K'}/8}+1, \qquad h=c'\log K'.
        \]
        To check \cref{eq:parameters}, $m(\alpha-1)\leq K'/4\leq K$ holds. For sufficiently large $K'$, $h_0\leq h\leq \alpha\leq n/2$ also holds, and 
        \[
            \frac{C_0\L(p)}{m}\leq (K')^{-c'} = 2^{-h}
        \]
        for an appropriately chosen $c'$, since $\L(p)/m=O(\delta/\sqrt{K'})$. In this case, $\Gamma=\Theta(\log K')=\Theta(h)$, therefore
        \[
            C  \gtrsim n \Gamma \log \frac{\sqrt{K'}}{\Gamma} \log\frac{n}{\Gamma}  - \frac{n}{\sqrt{K'}} \gtrsim n\Gamma \log \frac{K'}{\Gamma} \log \frac{n}{\Gamma}.
        \]
    \end{itemize}
\end{proof}


\bibliography{ref}

\clearpage

\appendix
\crefalias{section}{appendix}

\section{Proof of the randomized upper bound}
\label{app:rand-upper-proof}

We complete the proof of \cref{thm:rand-upper}.
Throughout this appendix, \(n\ge2\), \(K\ge1\), and \(c>0\) is fixed; constants hidden in \(O(\cdot)\) may depend only on \(c\).

\subsection{Outgoing-edge recovery}
\label[appendix]{subsec:rand-upper-recovery}

This subsection describes one recovery bundle and its query on a set \(C\subseteq V\).
We call the final cut \(\delta_G(C)\) \emph{small} if \(|\delta_G(C)|\le K\) and \emph{large} otherwise.
Small-cut copies use linear sketches, while historical forests handle large cuts.

Fix a bijection \(\id:\mathcal E\to[N]\), and identify edges with their indices when indexing vectors.
For each vertex \(v\), its signed incidence vector \(x^{(v)}\in\{-1,0,1\}^N\) has a \(+1\) at every final edge \(vu\) with \(v<u\), a \(-1\) at every final edge \(uv\) with \(u<v\), and zeros elsewhere.
An update \((e,\Delta)\), where \(e=uv\) and \(u<v\), changes coordinate \(e\) of \(x^{(u)}\) by \(\Delta\) and of \(x^{(v)}\) by \(-\Delta\).
Internal edges cancel when these vectors are summed over \(C\), so
\begin{equation}
    \supp\left(\sum_{v\in C}x^{(v)}\right)=\delta_G(C),
    \label{eq:rand-upper-cut-cancellation}
\end{equation}
and every nonzero coordinate of the sum is \(\pm1\).

Let \(p>2N\) be a prime, chosen in \cref{subsec:rand-upper-algorithm}.
A \emph{linear measurement} of \(y\in\mathbb Z^N\) is a value \(\sum_e\lambda_e y_e\in\mathbb F_p\) for fixed coefficients \(\lambda_e\in\mathbb F_p\).
A coordinate update by \(\Delta\) adds \(\lambda_e\Delta\) to the measurement.
Measurements with the same coefficients add across vertices, giving the measurement of \(\sum_{v\in C}x^{(v)}\).

\paragraph*{The singleton decoder.}
For \(x\in\{-1,0,1\}^N\) and \(\alpha\in\mathbb F_p\), define
\[
    m_1(x)\coloneqq\sum_e\id(e)x_e,
    \qquad
    m_2^\alpha(x)\coloneqq\sum_e\alpha^{\id(e)}x_e
    \quad\text{in }\mathbb F_p.
\]
Given these measurements, the decoder \(D_\alpha\) checks whether \(m_1(x)=\sigma j\) for some \(\sigma\in\{-1,+1\}\) and \(j\in[N]\).
Since \(p>2N\), such a pair is unique if it exists.
The decoder \emph{accepts} \(\id^{-1}(j)\) if also \(m_2^\alpha(x)=\sigma\alpha^j\), and \emph{rejects} otherwise.
The first measurement determines the candidate independently of \(\alpha\).

\begin{lemma}
\label{lem:rand-upper-decoder}
For every fixed \(x\in\{-1,0,1\}^N\), if \(|\supp(x)|=1\), then \(D_\alpha\) returns its unique edge for every \(\alpha\in\mathbb F_p\).
Otherwise, \(\Pr_\alpha[D_\alpha\text{ accepts}]\le N/p\) for uniform \(\alpha\in\mathbb F_p\).
\end{lemma}

\begin{proof}
The singleton case follows directly from the decoder's definition.
Suppose \(|\supp(x)|\ne1\).
The first measurement either yields no candidate, in which case the
decoder rejects, or fixes a pair \((\sigma,j)\) independently of
\(\alpha\).
In the latter case, acceptance is equivalent to \(P(\alpha)=0\), where
\[
    P(z)\coloneqq\sum_e x_e z^{\id(e)}-\sigma z^j
    \in\mathbb F_p[z].
\]
Since \(|\supp(x)|\ne1\) and \(p>2\), this polynomial is nonzero
and has degree at most \(N\).
The Schwartz--Zippel lemma therefore gives
\(\Pr_\alpha[D_\alpha\text{ accepts}]\le N/p\)
for uniform \(\alpha\in\mathbb F_p\).
\end{proof}

\paragraph*{Small-cut copies.}
Let \(L\coloneqq\lceil\log(2K)\rceil\) and \(b\coloneqq\max\{L,\lceil\log N\rceil\}\). Identify \([N]\) with a subset of \(\mathbb F_{2^b}\) and represent field elements as \(b\)-bit integers. Each \emph{small-cut copy} chooses independent uniform \(\beta,\beta'\in\mathbb F_{2^b}\) and an independent uniform fingerprint point \(\alpha\in\mathbb F_p\). The affine hash \(h(j)=\beta j+\beta'\) has pairwise-independent uniform outputs~\cite{carter_wegman:JCSS.1979}, since prescribed outputs at two distinct inputs determine the coefficients uniquely. The copy stores \(2b+\lceil\log_2 p\rceil\) bits for these choices.

The nested sampling levels are
\[
    S_\ell\coloneqq\{e\in\mathcal E:h(\id(e))<2^{b-\ell}\},
    \qquad \ell=1,\ldots,L.
\]
Since \(\ell\le b\), each edge belongs to \(S_\ell\) with probability \(2^{-\ell}\), pairwise independently at each fixed level.
For each vertex \(v\), the \emph{sketch} at level \(\ell\) consists of \(m_1\) and \(m_2^\alpha\) applied to \(x^{(v)}\) with all coordinates outside \(S_\ell\) set to zero.
Each copy uses the same \(h\) and \(\alpha\) for all vertices and levels.
Let \(x^{(C,\ell)}\) be the sum of these restricted vectors over \(v\in C\).
By linearity, summing the sketches over \(C\) gives the two measurements of \(x^{(C,\ell)}\), with
\begin{equation}
    \supp\left(x^{(C,\ell)}\right)=\delta_G(C)\cap S_\ell.
    \label{eq:rand-upper-level-support}
\end{equation}

\paragraph*{Historical random minimum spanning forests.}
A \emph{rank function} is an injection \(\pi:[N]\to\mathbb Z\); write \(\pi(e)=\pi(\id(e))\).
Distinct ranks give a unique minimum spanning forest, which contains the minimum-rank edge of every nonempty cut by the cut property.
We draw rank functions uniformly from a family of permutations of \([N]\) such that, for every nonempty \(I\subseteq[N]\) and \(j\in I\),
\begin{equation}
    \Pr_\pi\left[\pi(j)=\min_{i\in I}\pi(i)\right]
    \ge\frac{1-\varepsilon_0}{|I|},
    \label{eq:rand-upper-minwise}
\end{equation}
where \(\varepsilon_0\le1/2\).
Saks, Srinivasan, Zhou, and
Zuckerman~\cite{saks_et_al:IPL.2000} give, for
\(\varepsilon_0=N^{-\Theta(1)}\), an explicit family of
\(N^{O(\sqrt{\log N})}\) permutations satisfying
\eqref{eq:rand-upper-minwise}.
A permutation is specified by \(O(\log^{3/2}N)\) bits, and
\(\pi(j)\) can be computed from this specification and \(j\) in
time, and hence space, polylogarithmic in \(N\).
For the finitely many values of \(N\) for which this
\(\varepsilon_0\) exceeds \(1/2\), we store a uniformly random
permutation of \([N]\) explicitly.
It satisfies \eqref{eq:rand-upper-minwise} with
\(\varepsilon_0=0\) and uses \(O(1)\) bits.

For each sampled rank function \(\pi\), maintain an initially empty
forest \(T_\pi\) with one presence bit per stored edge.
On an insertion of \(e\), replace \(T_\pi\) by
\(T_\pi\cup\{e\}\) and, if a cycle is created, remove its
maximum-rank edge, which may be \(e\) itself.
Deletions leave \(T_\pi\) unchanged.
After each update to \(e\), set its presence bit to one for an
insertion and zero for a deletion whenever \(e\in T_\pi\), and
discard the bits of removed edges.

This insertion rule maintains \(T_\pi\) as the minimum spanning forest of the historical graph \(H\).
Reinserting a stored edge only resets its presence bit to one.
A reinserted edge outside \(T_\pi\) is the unique maximum-rank edge on the cycle it creates, so the update removes it again.
Each stored edge enters \(T_\pi\) on insertion, and its presence bit tracks subsequent updates while it remains stored.

\paragraph*{Recovery bundles and component queries.}
Fix \(q_s=40\) and \(q_m=16\).
A \emph{recovery bundle} consists of \(q_s\) small-cut copies and \(q_m\) historical forests with presence bits.
All hashes, fingerprint points, and rank functions are mutually independent.
A \emph{query} on \(C\) examines copies \(a=1,\ldots,q_s\) and, within each copy, levels \(\ell=1,\ldots,L\), in this order.
Each \emph{test} sums the vertex sketches and applies \(D_\alpha\); the query returns the first accepted edge.
If all tests reject, it scans the historical forests in a fixed order and returns the first present edge leaving \(C\), or \emph{fails} if none exists.
The same query applies to every component, since its cut size is unknown.

The \emph{ideal query} replaces each singleton decoder by an \emph{ideal decoder} that returns the unique edge of \(\supp(x^{(C,\ell)})\) if this support has size one, and rejects otherwise.
It depends on the hashes and rank functions, but not on the fingerprint points.

\begin{lemma}
\label{lem:rand-upper-recovery}
Consider one recovery bundle.
\begin{enumerate}
    \item[(a)] For every \(C\subseteq V\) and every outcome of its random choices, the ideal query either fails or returns an edge of \(\delta_G(C)\).
    In particular, it fails if \(\delta_G(C)=\varnothing\).
    \item[(b)] For every fixed \(C\subseteq V\) with \(\delta_G(C)\ne\varnothing\), the ideal query fails with probability at most
    \(\max\{(7/8)^{q_s},(3/4)^{q_m}\}<1/64\).
\end{enumerate}
\end{lemma}

\begin{proof}
Part (a) follows from~\eqref{eq:rand-upper-level-support} and the presence bits: both ideal singleton tests and forest scans return only present edges leaving \(C\).
For (b), write \(s_C=|\delta_G(C)|\).

If \(s_C\le K\), let \(\ell^*=\lceil\log (2s_C)\rceil\).
Then \(1\le\ell^*\le L\) and \(1/(4s_C)<2^{-\ell^*}\le1/(2s_C)\).
In one small-cut copy, pairwise independence and a union bound, summed over the disjoint events that a specified cut edge is the unique sampled edge, give
\[
    \Pr\left[|\delta_G(C)\cap S_{\ell^*}|=1\right]
    \ge s_C2^{-\ell^*}\bigl(1-(s_C-1)2^{-\ell^*}\bigr)
    >\frac18.
\]
Thus one copy succeeds at level \(\ell^*\) with probability at least \(1/8\), and the \(q_s\) independent copies fail with probability at most \((7/8)^{q_s}\).

If \(s_C>K\), at most \(K\) historical edges are absent, so \(|\delta_H(C)|\le s_C+K<2s_C\).
For one sampled rank function \(\pi\), let \(e_{\min}\) be the minimum-rank edge of \(\delta_H(C)\).
Summing~\eqref{eq:rand-upper-minwise} over the final cut edges gives
\[
    \Pr_\pi[e_{\min}\in\delta_G(C)]
    \ge(1-\varepsilon_0)\frac{s_C}{|\delta_H(C)|}
    >\frac14.
\]
On this event, \(e_{\min}\in T_\pi\) by the cut property and its presence bit is one, so the query succeeds.
The \(q_m\) independent forests therefore fail with probability at most \((3/4)^{q_m}\).
Both bounds are below \(1/64\) for the chosen constants.
\end{proof}

\begin{lemma}
\label{lem:rand-upper-bundle-space}
The state of one recovery bundle, including its random choices, can be stored in \(O(q_snL\log p+q_mn\log n)\) bits.
\end{lemma}

\begin{proof}
Each small-cut copy stores two elements of \(\mathbb F_p\) for each
of the \(n\) vertices and \(L\) levels, using
\(2nL\lceil\log p\rceil\) bits.
Its random choices use
\(2b+\lceil\log p\rceil=O(L+\log n+\log p)\) bits.
Since \(n\ge2\), \(L\ge1\), and \(\log p>\log n\), the total
is \(O(nL\log p)\) bits per copy.

Each historical forest stores at most \(n-1\) edges, each by its
index and a presence bit, using \(O(n\log n)\) bits.
Its rank function uses \(O(1+\log^{3/2}N)=O(n\log n)\) bits,
including the explicit representation for small \(N\).
Summing over \(q_s\) small-cut copies and \(q_m\) historical forests
gives \(O(q_snL\log p+q_mn\log n)\) bits.
\end{proof}

\subsection{The algorithm}
\label[appendix]{subsec:rand-upper-algorithm}
In our algorithm, in addition to \(L\), \(b\), \(q_s\), and \(q_m\), we also consider the number of rounds \(R\coloneqq\left\lceil\max\{10,3c\}\log (K+1)\right\rceil\) and a prime \(p\), chosen as follows.
In every round, the post-processing queries at most \(n\) sets, because the sets queried in a round partition \(V\), and every query performs at most \(q_sL\) tests.
Hence \(M_{\max}\coloneqq nq_sLR\) bounds the number of tests performed by the post-processing.
Before the streaming phase, the algorithm finds a prime \(p\) such that
\begin{equation}
    \max\{2N,\;M_{\max}Nn^c\}<p\le2\max\{2N,\;M_{\max}Nn^c\}.
    \label{eq:rand-upper-field-size}
\end{equation}
Such a prime exists by Bertrand's postulate.
The condition \(p>2N\) is needed by the decoder, and the condition \(p>M_{\max}Nn^c\) is used in \cref{lem:rand-upper-coupling}.
Since \(K\le N<n^2\), we have \(L=O(\log n)\), \(R=O(\log n)\), and \(M_{\max}=O(n\log^2n)\), so \(\log p=O(\log n)\).
All parameters are determined by \(n\), \(K=\min\{k,N\}\), and \(c\).

Before the stream, the algorithm samples and stores \(R\) mutually independent recovery bundles.
During the stream, it maintains \(F\) and its presence bits as in \cref{subsec:rand-upper-backbone} and updates every bundle as in \cref{subsec:rand-upper-recovery}.
After the stream, it runs the contraction procedure of \cref{subsec:rand-upper-overview} from \(\mathcal F=F_0\), using bundle \(r\) in round \(r\).
A disjoint-set structure maintains the components of \(\mathcal F\).
Let \(\mathcal C_r\) denote the partition of \(V\) into these components at the start of round \(r\).

\subsection{Progress of the contraction}
\label[appendix]{subsec:rand-upper-contraction}

Since the stream is oblivious, we fix it for the rest of the analysis.
All probabilities are over the random choices of the algorithm.
The \emph{ideal process} is the post-processing of
\cref{subsec:rand-upper-algorithm} with every query replaced by the
corresponding ideal query.
Its execution depends on the hash functions and rank functions of the
bundles, but not on the fingerprint points.
In this subsection, \(\mathcal F\), \(\mathcal C_r\), and the candidates
refer to the ideal process.

For \(r=0,\ldots,R\), let \(\mathcal G_r\) denote the hash functions
and rank functions of bundles \(1,\ldots,r\).
The execution of the first \(r\) rounds is determined by \(\mathcal G_r\).
For \(r=1,\ldots,R+1\), let \(A_r\) be the number of active components
at the beginning of round \(r\), where round \(R+1\) refers to the output.
For \(r\in[R]\), let \(Y_r\) be the number of active components
that obtain no candidate in round \(r\).

\begin{lemma}
\label{lem:rand-upper-one-round}
In every round \(r\in[R]\) of the ideal process,
\(\mathcal F\) remains a forest contained in \(E\), and \(A_{r+1}\le (A_r+Y_r)/2\).
\end{lemma}

\begin{proof}
By \cref{lem:rand-upper-recovery}(a), every candidate of a component
\(C\in\mathcal C_r\) lies in \(\delta_G(C)\).
Since the process starts from \(F_0\subseteq E\) and adds an edge only
if it joins two distinct current components, \(\mathcal F\) remains
a forest contained in \(E\).
An initially inactive component has no outgoing edge in \(G\),
so it never merges and remains inactive.

The endpoints of every candidate are connected after the round,
whether the candidate is added or skipped.
Thus, every component that obtains a candidate belongs to a final
component containing at least two components of \(\mathcal C_r\).
Consequently, at most \(Y_r\) final active components contain only one
component of \(\mathcal C_r\); every other final active component
contains at least two, all initially active.
Together, they contain at least \(2A_{r+1}-Y_r\) initially active components.
Hence \(2A_{r+1}-Y_r\le A_r\), proving the bound.
\end{proof}

We now bound the expected number of active components after each round.

\begin{lemma}
\label{lem:rand-upper-convergence}
For any number \(R\ge1\) of rounds, the ideal process ends with an
active component with probability at most \(2K(65/128)^R\).
For the choice of \(R\) in \cref{subsec:rand-upper-algorithm},
this probability is at most \((K+1)^{-c}\).
\end{lemma}

\begin{proof}
Condition on \(\mathcal G_{r-1}\).
This fixes the partition \(\mathcal C_r\), while the hash functions
and rank functions of bundle \(r\) retain their distribution.
By \cref{lem:rand-upper-recovery}(b), each active component obtains
no candidate with conditional probability at most \(1/64\).
Linearity of expectation therefore gives \(\mathbb E[Y_r\mid\mathcal G_{r-1}]\le A_r/64\).
By \cref{lem:rand-upper-recovery}(a), every candidate is a valid
outgoing edge, so \cref{lem:rand-upper-one-round} yields
\[
    \mathbb E[A_{r+1}\mid\mathcal G_{r-1}]
    \le\frac{A_r+\mathbb E[Y_r\mid\mathcal G_{r-1}]}{2}
    \le\frac{65}{128}A_r.
\]
Iterating this bound and using \(A_1\le2K\) from
\cref{lem:rand-upper-backbone}, we obtain
\[
    \Pr[A_{R+1}>0]
    \le\mathbb E[A_{R+1}]
    \le2K\left(\frac{65}{128}\right)^R,
\]
where the first inequality follows from Markov's inequality.
Since \(65/128\le 2^{-2/3}\) and \(\frac{2}{3}\max \{10, 3c\} \ge c+2\), our choice of \(R\) gives
\[
2K(65/128)^R\le 2K(K+1)^{-(c+2)}\le (K+1)^{-c},
\]
where the last inequality uses \(2K\le (K+1)^2\).
\end{proof}

\subsection{Fingerprints, correctness, and space}
\label[appendix]{subsec:rand-upper-completion}

We now bound the fingerprint error and compare the algorithm with the ideal process.
For each round \(r\in[R]\) and component \(C\in\mathcal C_r\) of the ideal process, consider the tests at all \(L\) levels of all \(q_s\) small-cut copies of bundle \(r\), including tests skipped after an earlier acceptance.
We call these \emph{potential tests}.
There are at most \(M_{\max}=nq_sLR\), since \(|\mathcal C_r|\le n\).
Let \(\mathcal B_{\mathrm{fp}}\) be the event that the actual decoder accepts a vector with support size different from one in some potential test.

\begin{lemma}
\label{lem:rand-upper-coupling}
\(\Pr[\mathcal B_{\mathrm{fp}}]\le M_{\max}N/p<n^{-c}\).
If \(\mathcal B_{\mathrm{fp}}\) does not occur, then in every round the algorithm and the ideal process have the same partition and record the same candidates, and they produce the same output.
\end{lemma}

\begin{proof}
Condition on \(\mathcal G_R\), which fixes the ideal process and the vector of every potential test, while each fingerprint point remains uniform.
By \cref{lem:rand-upper-decoder} and a union bound, \(\Pr[\mathcal B_{\mathrm{fp}}\mid\mathcal G_R]\le M_{\max}N/p\).
Averaging over \(\mathcal G_R\) and using~\eqref{eq:rand-upper-field-size} gives \(\Pr[\mathcal B_{\mathrm{fp}}]\le M_{\max}\frac{N}{p}<n^{-c}\).

Suppose that \(\mathcal B_{\mathrm{fp}}\) does not occur.
Then both decoders give the same answer on every potential test by \cref{lem:rand-upper-decoder}.
Both processes start from \(F_0\).
If they begin a round with the same partition, each component query encounters the same tests in the same order, with the same answers, and either returns the same first accepted edge or reaches the identical forest scan.
Thus both processes record the same candidates and perform the same merges.
Induction over the rounds shows that their partitions, candidates, and outputs agree.
\end{proof}

\begin{proof}[Proof of \cref{thm:rand-upper}]
\textit{Correctness.}
The ideal process starts from \(F_0\subseteq E\) and remains a forest contained in \(E\) by \cref{lem:rand-upper-recovery}(a) and \cref{lem:rand-upper-one-round}.
By \cref{lem:rand-upper-convergence,lem:rand-upper-backbone}, its output is a spanning forest of \(G\) except with probability at most \((K+1)^{-c}\).
By \cref{lem:rand-upper-coupling}, the algorithm has the same output except with probability at most \(n^{-c}\).
A union bound gives the claimed error \((K+1)^{-c}+n^{-c}\).

\textit{Space.}
The following bounds hold for every stream and every outcome of the random choices.
\begin{itemize}
    \item The forest \(F\) and its presence bits use \(O(n\log n)\) bits.
    \item By \cref{lem:rand-upper-bundle-space}, each bundle, including its random choices, uses \(O(n\log n\log K)\) bits, since \(q_s,q_m=O(1)\), \(L=O(\log K)\), and \(\log p=O(\log n)\).
    The \(R=O(\log K)\) bundles therefore use \(O(n\log n\log^2K)\) bits.
    \item The parameters and field representation use \(O(\log n)\) bits.
    During the stream, \(O(n\log n)\) bits of reusable workspace suffice for connectivity and maximum-rank path queries by depth-first search.
    This bound also covers rank evaluation and repeated squaring.
    After the stream, the disjoint-set structure, output forest, and at most \(n\) candidates use \(O(n\log n)\) bits, with \(O(\log p)\) additional bits for sketch aggregation, decoding, and forest scans.
\end{itemize}
The total is \(O(n\log n\log^2K)\) bits, with a hidden constant depending only on \(c\).
\end{proof}

For \(K=n^{o(1)}\), the bound in \cref{thm:rand-upper} does not guarantee inverse-polynomial failure probability in \(n\). In \cref{sec:high-prob-upper}, we obtain this guarantee using a sparse-recovery sketch shared across rounds.

\section{Optimal Deterministic Algorithm}
\label{sec:kec-deter}

We determine the deterministic single-pass space complexity of
computing a spanning tree in the bounded-deletion model. Our algorithm returns a spanning tree if the final graph is connected and an edge-maximal spanning forest otherwise. Our lower bounds hold even when the final graph is promised to be connected.

\begin{theorem}\label{thm:deter-forest}
For every integer $k\ge1$, the deterministic single-pass space
complexity of computing a spanning tree in an $n$-vertex graph
stream with at most $k$ deletions is
\[
 \Theta\left(\min\left\{n^2,\,
 nk\log\left(1+\frac nk\right)\right\}\right)
 \text{ bits},
\]
where the deletion bound $k$ is given in advance. The lower
bound holds even for connected final graphs, and the upper
bound also applies to computing an edge-maximal spanning forest
in arbitrary final graphs.
\end{theorem}

We prove the upper bound in \cref{sec:deter-algo} and the lower bound in
\cref{sec:deter-lower}.

\subsection{Algorithm}
\label{sec:deter-algo}

If $k=\Omega(n)$, then $nk\log(1+n/k)=\Theta(n^2)$, and
storing the entire graph as an adjacency matrix gives the desired
$O(n^2)$-bit upper bound. Hence, in what follows, we assume
$k=o(n)$.

Let $I$ and $D$ be the multisets
of inserted and deleted edges, respectively, so that the final
graph is $G=(V,I\setminus D)$. Multisets are necessary because an
edge may be deleted and later reinserted. Throughout this subsection,
multiset unions add multiplicities, and multiset subtraction
truncates negative multiplicities to zero.

For a multigraph $X$, let $\lambda_X(s,t)$ denote the maximum
number of edge-disjoint $s$--$t$ paths in $X$. A spanning subgraph
$A\subseteq X$ is a \emph{$q$-edge-connectivity certificate} of $X$ if
\[
 \lambda_A(s,t)\ge \min\{q,\lambda_X(s,t)\}
 \qquad\text{for all distinct }s,t\in V.
\]
We identify an edge multiset with its multigraph on $V$.
We maintain a $(k+1)$-edge-connectivity certificate $A$ of the
insertion multiset $I$, together with the multiset $B$ of all
deletions. At the end of the stream, $B=D$, and we return an
edge-maximal spanning forest of $(V,A\setminus B)$.

\begin{lemma}\label{lem:prefix-cut}
If $A$ is a $(k+1)$-edge-connectivity certificate of $I$ and
$|D|\le k$, then $(V,A\setminus D)$ and $G=(V,I\setminus D)$
have the same connected components.
\end{lemma}

\begin{proof}
Since $A\setminus D\subseteq I\setminus D$, it suffices to rule
out a cut with no crossing edge in $A\setminus D$ but a crossing
edge $\{s,t\}$ in $G$. Suppose such a cut exists. All edges of $A$
crossing this cut are removed by $D$, so there are at most $k$
of them. Hence $\lambda_A(s,t)\le k$.

Moreover, some copy of $\{s,t\}$ in $I$ is absent from $A$; otherwise
a surviving copy would also belong to $A\setminus D$. Adding
this copy to a maximum collection of edge-disjoint $s$--$t$
paths in $A$ gives
$\lambda_I(s,t)\ge\lambda_A(s,t)+1$. The certificate property
therefore implies
\[
 \lambda_A(s,t)
 \ge \min\{k+1,\lambda_I(s,t)\}
 \ge \lambda_A(s,t)+1,
\]
a contradiction.
\end{proof}

By the standard sparse-certificate construction, the union of
$k+1$ successive residual edge-maximal spanning forests is a
$(k+1)$-edge-connectivity certificate with at most $(k+1)(n-1)$
edges~\cite{nagamochi1992linear}. Using the strong-certificate
composition principle of Eppstein et al.~\cite{EppsteinGIN97},
we maintain such a certificate $A$ of the insertion multiset:
initially $A=B=\emptyset$; upon inserting $e$, replace $A$ by
a sparse certificate of $A\cup\{e\}$; upon deleting $e$, add
one copy of $e$ to $B$. Thus $A$ always certifies the insertions
processed so far, and \cref{lem:prefix-cut} proves correctness.

It remains to encode $A$ without storing a forest index for
each edge. Rank the possible unordered vertex pairs in a
universe of size $N=\binom n2$. If $A$ has $m\ge1$ distinct edges, counted without multiplicity, list their ranks as $x_1\le\cdots\le x_m$ and encode the strictly
increasing sequence $x_i$
using Elias--Fano
encoding~\cite{elias1974efficient}.  This takes
$O(m\log(2+N/m))$ bits. There are at most $k$ copies beyond the first copy of each edge, since every reinsertion requires a preceding deletion. Storing one edge ID for each extra copy takes an additional
$O(k\log n)$ bits. Since $m\le(k+1)(n-1)=O(nk)$ and $k=o(n)$,
the space is
\(
 O\left(nk\log(1+n/k)\right)\text{ bits}.
\)
The deletion multiset $B$ uses $O(k\log n)$ additional bits.
Certificate reconstruction can process one residual forest
at a time, using compressed residual and output multisets
and $O(n\log n)$ auxiliary bits. The final spanning forest can
be extracted within the same space bound.

\subsection{Lower Bound}
\label{sec:deter-lower}

Our proof uses a reconstruction argument similar to that of Khanna, Konrad, and Dark~\cite{khanna_konrad_dark:ICALP.2025} for deterministic maximal matching in bounded-deletion streams. We encode a regular bipartite graph in the memory state of a
deterministic streaming algorithm. A fixed auxiliary vertex
ensures connectedness, while repeated executions from the same
state with different deletion sets reveal the entire encoded graph.

For sufficiently large $n$, set
\[
 r=\min\left\{k+1,\left\lfloor\frac{n-1}{4}\right\rfloor\right\},
 \qquad
 t=\left\lfloor\frac{n-1}{2r}\right\rfloor.
\]
Take disjoint vertex sets $X$ and $Y$, each partitioned into
$r$ groups of size $t$. For every pair of groups, one in $X$
and one in $Y$, choose an arbitrary perfect matching between
them. Their union is a simple $r$-regular bipartite graph $Q$.
Let $\mathcal Q$ denote the family of graphs obtained this way.
The independent choices give
\(
 |\mathcal Q|=(t!)^{r^2}.
\)

Add a new vertex $z$ adjacent to every vertex of $Y$, and attach
the remaining $n-1-2rt$ vertices as leaves to $z$. Alice inserts
the resulting connected graph in a fixed edge order and sends
the algorithm's memory state to Bob. We show that this state
uniquely determines $Q$.

Fix $x\in X$, and initialize $S=\emptyset$. Bob takes a fresh
copy of Alice's state, deletes the edges in $S$, and obtains an
output spanning tree. As long as $|S|\le r-1$, at least one of
the $r$ edges incident to $x$ survives. The final graph remains
connected: $z$ connects all vertices of $Y$, every vertex of $X$
retains a neighbor in $Y$, and the added leaves remain attached.
The output tree therefore contains an edge incident to $x$
outside $S$. Bob adds one such edge to $S$ and repeats until
all $r$ incident edges have been recovered. Each execution uses
at most $r-1\le k$ deletions. Repeating this procedure for every
$x\in X$ recovers $Q$ completely.

Consequently, distinct graphs in $\mathcal Q$ require distinct
memory states. If the algorithm uses $b$ bits, then
\[
 b\ge\log|\mathcal Q|
 =r^2\log(t!)
 =\Omega\left(nr\log\left(1+\frac nr\right)\right)
 =\Omega\left(\min\left\{n^2,\,
 nk\log\left(1+\frac nk\right)\right\}\right),
\]
where the simplifications follow from $t\ge2$, $rt=\Theta(n)$,
$r=\Theta(\min\{k,n\})$, and
$\log(t!)=\Theta(t\log t)$.
This proves the lower bound in \cref{thm:deter-forest}.
Every graph queried during decoding is connected, so the bound
holds even under the connectedness promise.

\subsection{Edge-Connectivity Certificates}
\label{sec:deter-cert}

The same approach extends to computing a
$\lambda$-edge-connectivity certificate. If $k+\lambda\ge n/4$, we maintain the graph in an adjacency matrix using $O(n^2)$ bits and extract the certificate at the end of the stream. Otherwise, as in \cref{sec:deter-algo}, we maintain a $(k+\lambda)$-edge-connectivity certificate of the insertion multiset, using $k+\lambda$ successive residual spanning forests at each sparsification step. Every cut is either preserved completely
or contains at least $k+\lambda$ certificate edges. After at
most $k$ deletions, the former cuts remain exact and the latter
retain at least $\lambda$ edges. Thus subtracting the deletion
multiset leaves a $\lambda$-edge-connectivity certificate of the
final graph. Taking $\lambda$ successive residual spanning
forests reduces the output to at most $\lambda(n-1)$ edges.
The same encoding argument gives the upper bound below.

For the lower bound, use the construction of \cref{sec:deter-lower}
with 
\[
 r=\min\left\{k+\lambda,\left\lfloor\frac{n-1}{4}\right\rfloor\right\}.
\]
For each $x\in X$, Bob recovers $\max\{0,r-\lambda\}$ incident edges one at a time. In each execution, he starts from a fresh copy of
Alice's memory state, deletes the edges of $x$ already recovered, and records one incident edge from the output certificate. In a final execution, Bob deletes all the recovered edges of $x$. The degree of $x$ is then $\min\{r,\lambda\}\le\lambda$,
so the certificate must contain every remaining edge incident to $x$. Each execution uses at most $\max\{0,r-\lambda\}\le k$ deletions. Since at least one
edge incident to $x$ survives, the graph remains connected, as in \cref{sec:deter-lower}. Repeating for each $x\in X$ recovers the entire bipartite graph. The same counting argument gives a lower bound of $\Omega\bigl(nr\log(1+n/r)\bigr)$ bits. Substituting the choice of $r$ gives the stated bound.

\begin{corollary}\label{thm:lambda-certificate}
For integers $k\ge1$ and $\lambda\ge1$, the deterministic
single-pass space complexity of computing a
$\lambda$-edge-connectivity certificate with at most
$\lambda(n-1)$ edges in an $n$-vertex graph stream with at most
$k$ deletions is
\[
 \Theta\left(\min\left\{n^2,\,
 n(k+\lambda)\log\left(1+\frac{n}{k+\lambda}\right)
 \right\}\right)\text{ bits},
\]
where $k$ and $\lambda$ are given in advance. The lower bound
holds even when the final graph is promised to be connected.
\end{corollary}

\section{Preliminaries on Information Complexity} \label{sec:IC}
In this section, we present the preliminaries of information complexity relevant to our analysis and refer the reader to standard references such as \cite{RY20,Bra19,CT05} for a comprehensive introduction to the subject.

\paragraph*{Entropy and Mutual Information}
When \(\mu\) is a joint distribution over several random variables, we occasionally write \(\mu[XY]\) for the marginal distribution on \((X,Y)\) instead of \(\mu_{XY}\) to avoid cumbersome subscripts.
For random variables $X,Y,Z$ distributed according to the joint distribution $\mu$, we denote $\H_{\mu}(X)$ and $\H_{\mu}(X\mid Y)$ the (Shannon) entropy of $X$ and the conditional entropy of $X$ conditioned on $Y$ respectively, and we drop the subscript $\mu$ whenever the context is clear. Similarly, $\I_\mu(X;Y)\defeq \H_\mu(X) - \H_\mu(X\mid Y)$ denotes the mutual information between $X$ and $Y$, and $\I_\mu(X;Y\mid Z)\defeq \H_\mu(X\mid Z) - \H_\mu(X\mid YZ)$ denotes the conditional mutual information between $X$ and $Y$ conditioned on $Z$.

\begin{fact}[Non-negativity of Entropy and Mutual information]
    $\H(\cdot)$ and $\I(\cdot;\cdot)$ are non-negative functions.
\end{fact}

\begin{fact}[Chain Rule and Sub-additivity for Entropy]
    For random variables $X_1,\ldots,X_n$, 
    \[
        \H(X_1\ldots X_n) = \sum_{i=1}^n \H(X_i\mid X_{<i})
    \]
    where $X_{<i}\defeq X_1\ldots X_{i-1}$. Since conditioning cannot increase entropy, 
    \[
        \H(X_1\ldots X_n) \leq \sum_{i=1}^n \H(X_i).
    \]
\end{fact}

\begin{fact}[Chain Rule for Mutual Information]
    For random variables $X_1,\ldots,X_n,Y,Z$, 
    \[
        \I(X_1\ldots X_n;Y\mid Z) = \sum_{i=1}^n \I(X_i;Y\mid X_{<i}Z)
    \]
\end{fact}

\begin{fact}[Independent event and mutual information]
    For random variables $X,Y,Z,W$, if $W$ is independent from $X,Y,Z$, then $\I(X;Y\mid ZW) = \I(X;Y\mid Z)$.
\end{fact}

\begin{fact}[Data processing inequality for mutual information]\label{fact:DPI-MI}
    For every random variables $X,Y,Z$ and function $f$, 
    \[\I(X; f(Y)\mid Z) \leq \I(X; Y\mid Z).\]
\end{fact}

The following lemma is useful for our discussion of one-way communication protocols.
\begin{lemma}\label{lem:one-way}
    Let $\Pi$ be a randomized one-way protocol
    with Alice's input $A$, Bob's input $B$, public randomness $R$, and Alice's message $M$.
    For any joint distribution of $(A,B)$,
    \[
        \I(A;M\mid BR)\leq \I(A;M\mid R).
    \]
\end{lemma}

\begin{proof}
    Since $\Pi$ is one-way, Alice's message $M$ is conditionally independent of $B$ given $(A,R)$. Thus,
    $\I(B;M\mid A,R)=0$.
    By the chain rule for mutual information,
    \[ \I(A;M\mid R)
        = \I(AB;M\mid R)
        = \I(B;M\mid R)+\I(A;M\mid BR)
        \geq \I(A;M\mid BR). \qedhere
    \]
\end{proof}

\paragraph*{Kullback--Leibler (KL) divergence}
For two probability distributions $P$ and $Q$ over the same
discrete space $\Omega$, the \emph{KL divergence} is defined as
\[
    \KL(P\Vert Q)
    = \sum_{x\in\Omega} P(x)\log\frac{P(x)}{Q(x)}.
\]
In particular, $\KL(P\Vert Q)<\infty$ implies that $\supp(P)\subseteq \supp(Q)$.
\begin{fact}[Mutual information defined by KL divergence]\label{fact:KL-MI}
    For every discrete random variables $X$ and $Y$,
    \[
        \I(X;Y)
        = \KL\left(\doublefrac{P[XY]}{P[X]P[Y]}\right)
        = \E_{x\sim P_X}
          \left[\KL\left(\doublefrac{P[Y\mid X=x]}{P[Y]}\right)\right].
    \]
\end{fact}
\begin{fact}[Pinsker's inequality]
    Let $P$ and $Q$ be discrete distributions over a discrete domain $\Omega$. Denote $\TV(P,Q)$ the \emph{total variation distance} between $P$ and $Q$, defined as
    \[
        \TV(P,Q) = \frac{1}{2}\sum_{x\in\Omega} \abs{P(x)-Q(x)} = \sum_{x\in \Omega} \max\set{P(x)-Q(x),0}
            = \max_{S\subseteq\Omega}\abs{P(S)-Q(S)}.
    \]
    Then $\TV(P,Q)
    \leq \sqrt{\frac{\ln 2}{2} \KL(P\Vert Q)}$. 
\end{fact}
For conditional distributions $P[Y\mid X]$ and $Q[Y\mid X]$, the \emph{conditional relative entropy} with respect to the marginal $P[X]$ is defined as
\[
    \CRE{P[Y\mid X]}{Q[Y\mid X]}{P[X]}
    \defeq \E_{x\sim P[X]} \left[\KL\left(\doublefrac{P[Y\mid X=x]}{Q[Y\mid X=x]}\right)\right].
\]
\begin{fact}[Chain Rule for KL Divergence]\label{fact:KL-chain}
Let $P$ and $Q$ be two joint distributions of $(X,Y)$. Then
\[
    \KL(P[XY]\Vert Q[XY]) = \KL(P[X]\Vert Q[X]) + \CRE{P[Y\mid X]}{Q[Y\mid X]}{P[X]}.
\]
\end{fact}
\begin{fact}[Data processing inequality for KL divergence]\label{fact:DPI-KL}
    For every random variables $X,Y$ and function $f$, 
    \[\KL(f(X)\Vert f(Y)) \leq \KL(X\Vert Y).\]
\end{fact}

The following lemma appears as the second part of \cite[Lemma 19]{Braverman_et_al:FOCS.2013}, which is quoted by Nelson and Yu as Lemma 4 \cite{nelson_yu:SODA.2019}. 
\begin{lemma}[\cite{Braverman_et_al:FOCS.2013}]\label{lem:cond-ind-MI-bound}
    Let $\scW$ be an event and $A,B,R$ be random variables. Suppose $P$ is a distribution and $Q=P(\cdot\mid \scW)$. If $A$ and $B$ conditionally independent given $R$ under $P$, then 
    \[
        \I_Q(A;B\mid R)\leq \L(P(\scW)).
    \]
\end{lemma}
We provide a self-contained proof using properties of KL divergence.
\begin{proof}
    By the definition of $Q=P(\cdot \mid \scW)$, $\KL(Q\Vert P) = \L(P(\scW))$. 
    By the conditional independence given $R$,
    \[
        P_{ABR}(a,b,r) = P_R(r)P_{A\mid r} (a)P_{B\mid r}(b).
    \]
    Note that 
    \begin{align*}        
        \log\frac{Q_{ABR}(a,b,r)}{P_{ABR}(a,b,r)}
        &=\log \frac{Q_R(r)Q_{AB\mid r}(a,b)}{P_R(r) P_{A\mid r}(a)P_{B\mid r}(b)}\\
        &=\log \frac{Q_R(r)}{P_R(r)} +\log\frac{Q_{AB\mid r}(a,b)}{P_{A\mid r}(a)P_{B\mid r}(b)}\\
        &=\log \frac{Q_R(r)}{P_R(r)} +\log\frac{Q_{AB\mid r}(a,b)}{Q_{A\mid r}(a)Q_{B\mid r}(b)} + \log\frac{Q_{A\mid r}(a)Q_{B\mid r}(b)}{P_{A\mid r}(a)P_{B\mid r}(b)}\\
        &=\log \frac{Q_R(r)}{P_R(r)} +\log\frac{Q_{AB\mid r}(a,b)}{Q_{A\mid r}(a)Q_{B\mid r}(b)} + \log\frac{Q_{A\mid r}(a)}{P_{A\mid r}(a)} +\log\frac{Q_{B\mid r}(b)}{P_{B\mid r}(b)}.
    \end{align*}
    Taking $\E_Q$, the left-hand side becomes $\KL(Q_{ABR}\Vert P_{ABR})$, and the last sum becomes
    \begin{equation}\label{eq:cond-ind}
        \KL(Q_R\Vert P_R) + \I_Q(A;B\mid R) + \CRE{Q_{A\mid R}}{P_{A\mid R}}{Q_R} + \CRE{Q_{B\mid R}}{P_{B\mid R}}{Q_R}. 
    \end{equation}
    By non-negativity and data processing inequality,
    \[
        \I_Q(A;B\mid R)\leq \KL(Q_{ABR}\Vert P_{ABR})\leq \KL(Q\Vert P) = \L(P(\scW)). \qedhere
    \]
\end{proof}
\section{Proof of \cref{lem:direct-product}} \label{sec:direct-product}
We first begin with a proof overview. Let $i\in [m]$, $S,T\subseteq [m]$ such that $i\notin S\cup T$, and $\vec{r}$ be a possible assignment to $(\vec{X}_S,\vec{Y}_T)$. Here, $\vec{X}_S$ denotes the restriction $(\vec{X}_j)_{j\in S}$ and $\vec{Y}_T$ denotes the restriction $(\vec{Y}_j)_{j\in T}$.

In the first step, we construct a single-instance distribution and a lossy protocol for every \emph{configuration} $c=(i,S,T,\vec{r})$ conditioned on simultaneous success of $\Pi$ on computing $f^{\otimes m}$.

Next, we consider two equivalent ways of sampling $c$. The first view produces a forward chain-rule sum over Alice's inputs, and the second view produces a backward chain-rule sum over Bob's inputs.

Throughout the analysis, we control six quantities in expectations over configurations. By a union bound and Markov's inequality, there is a fixing of the configuration that satisfies all six bounds. The instance distribution and lossy protocol given by this configuration fulfill all requirements in the lemma.

\subsection{Input Distribution and Lossy Protocol Constructions}\label{subsec:input-distr}
Let $P$ be the joint distribution induced by $\mu^{\otimes m}$ and $\Pi$, and $M=(R_\Pi,M_\Pi)$. Since $R_\Pi$ is independent of the input pair $(X,Y)$ under $P$, by mutual information chain rule 
\begin{equation}\label{eq:MI-bound}
    \I_P(\vec{X};M\mid \vec{Y}) = \I_P(\vec{X};M_\Pi \mid YR_\Pi)\leq \gamma.
\end{equation}
Denote $\vec{z} = (z_1,\ldots,z_m)$ Bob's output vector. Then for every choice $(\vec{x},\vec{y},\tau,\vec{z})$,
\begin{equation*}
    P(\vec{x},\vec{y},\tau,\vec{z})=\mu^{\otimes m}(\vec{x},\vec{y})\cdot P(M=\tau\mid \vec{X}=\vec{x})\cdot P(\vec{Z}=\vec{z}\mid M=\tau,\vec{Y}=\vec{y}).
\end{equation*}
Define $\scW$ the event that $f(X_j,Y_j)=Z_j$ for every $j\in [m]$, and $Q=P(\cdot\mid \scW)$. In this notation, $P(\scW)\geq p$ and $\KL(Q\Vert P)=\L(P(\scW)) \leq \L(p)$.

For $i\in [m]$ and $S,T\subseteq[m]$, let $\vec{r}$ be an assignment to $(\vec{X}_S,\vec{Y}_T)$ with positive probability under $Q$. Write $c=(i,S,T,\vec{r})$ and define
\[
    \mu'_{c} = Q[X_iY_i\mid (\vec{X}_S,\vec{Y}_T)=\vec{r}]
    \quad\text{and}\quad
    q_{c} = Q[X_iY_iMZ_i\mid (\vec{X}_S,\vec{Y}_T)=\vec{r}].
\]
The lossy protocol $\Pi_{c}$ uses $q_{c}(M\mid X_iY_i)$ as the message generator $\Msg_{c}$, and $q_c(Z_i\mid MY_i)$ as the decoder $\Dec_c$. Under the input distribution $\mu'_c$, $\Pi_{c}$ has the joint distribution
\[
    \rho_c(x,y,\tau,z) = q_c(x,y)\Msg_{c}(\tau\mid x,y)\Dec_c(z\mid \tau,y) = q_c(x,y,\tau)q_c(z\mid \tau,y).
\]
In particular, $\rho_c$ and $q_c$ share the same $(X,Y,M)$ marginal, while $\Dec_c$ only depends on Bob's available information $(Y,M)$.

\def\FC{\mathsf{FC}}
\def\BC{\mathsf{BC}}
\def\RD{\mathsf{RD}}
For the configuration $c=(i,S,T,\vec{r})$, define 
\begin{align*}
    \FC_c &\defeq \I_Q(X_i;M\mid Y_i,(\vec{X}_S,\vec{Y}_T)=\vec{r}) = \I_{\rho_c}(X;M\mid Y)\\
    \BC_c &\defeq \I_Q(Y_i;M\mid X_i,(\vec{X}_S,\vec{Y}_T)=\vec{r}) = \I_{\rho_c}(Y;M\mid X)\\
    \RD_c &\defeq \I_Q(X_i;Z_i\mid MY_i,(\vec{X}_S,\vec{Y}_T)=\vec{r}) = \I_{q_c}(X;Z\mid MY)
\end{align*}
$\FC_c$ and $\BC_c$ represent the forward and backward information costs of the lossy protocol respectively, and $\RD_c$ is the residue dependence of the output on Alice's input. By KL divergence chain rule (\cref{fact:KL-chain}) and KL divergence definition of mutual information (\cref{fact:KL-MI}),
\begin{align*}
    \KL(q_c\Vert \rho_c) 
    &= \sum_{x,y,\tau,z} q_c(x,y,\tau,z)\log\frac{q_c(x,y,\tau,z)}{\rho_c(x,y,\tau,z)} \\
    &= \sum_{x,y,\tau,z} q_c(x,y,\tau,z)\log\frac{q_c(z\mid x,y,\tau)}{q_c(z\mid y,\tau)}\\
    &= \sum_{x,y,\tau} q_c(x,y,\tau) \sum_z q_c(z\mid x,y,\tau) \log\frac{q_c(z\mid x,y,\tau)}{q_c(z\mid y,\tau)}\\
    &= \E_{xy\tau\sim q_c[XYM]} \KL\left(\doublefrac{q_c(Z\mid x,y,\tau)}{q_c(Z\mid y,\tau)}\right)\\
    &= \I_{q_c}(X;Z\mid MY) \\
    &= \RD_c.
\end{align*}

Let $\scA=\set{(x,y,\tau,z): f(x,y)=z}$ be the event that the selected instance is answered correctly. Since $q_c$ is obtained by conditioning $\Pi$ on the event that all $m$ instances are answered correctly, we have
$q_c(\scA)=1$. On the other hand, $\Dec_c$ may make errors. Let $e_c=\Pr_{\rho_c}[f(X,Y)\neq Z]=1-\rho_c(\scA)$. Applying data processing inequality with the indicator of $\scA$ (\cref{fact:DPI-KL}), 
\[
    \KL(q_c\Vert \rho_c) \geq \KL(\1_{\scA}\circ q_c\Vert \1_{\scA}\circ \rho_c) = \KL(\textrm{Bern}(1)\Vert \textrm{Bern}(1-e_c))=\L(1-e_c), 
\]
implying that
\begin{equation}\label{eq:err-bound}
    e_c\leq 1-2^{-\RD_c}\leq (\ln 2) \RD_c.
\end{equation}

\subsection{Configuration Sampling}\label{subsec:config-sampling}
A configuration $c=(i,S,T,\vec{r})$ can be sampled in one of two equivalent ways. Independently choose a uniform permutation $\sigma$ over $[m]$, a uniform index $\lambda_S\in [m/2+1,m]$ and a uniform index $\lambda_T\in [1,m/2]$. Then
\begin{itemize}
    \item \emph{Forward view}: $i=\sigma(\lambda_S)$, $S=\sigma([1,\lambda_S-1])$, $T=\sigma([\lambda_T,m])\setminus\set{i}$;
    \item \emph{Backward view}: $i=\sigma(\lambda_T)$, $S=\sigma([1,\lambda_S])\setminus\set{i}$, $T=\sigma([\lambda_T+1,m])$.
\end{itemize}
The triple $(i,S,T)$ is identically distributed under both distributions.
Given $(i,S,T)$, $\vec{r}$ is sampled from $Q[\vec{X}_S\vec{Y}_T]$. It is clear that $i\notin S\cup T$ since $S\cup T=[m]\setminus \set{i}$. Two views allow us to derive an upper bound for expected forward and backward information costs respectively.

In the remaining calculations, $\E_c$ denotes this configuration distribution. Given a permutation $\sigma$, we write $X^\sigma_\lambda = X_{\sigma(\lambda)}$ and $\vec{X}^{\sigma}_{\leq \lambda}$ for $\vec{X}_{\sigma([1,\ldots,\lambda])}$, and extend the notations to other types of inequality signs.

\paragraph*{Forward view.}
In the forward view, $Y_i\vec{Y}_T = \vec{Y}^\sigma_{\geq \lambda_T}$, therefore by chain rule,
\begin{align}
    \E_c [\FC_c]
    &= \E_{\sigma,\lambda_S,\lambda_T} \I_Q\left(X^\sigma_{\lambda_S};M\mid \vec{X}^\sigma_{<\lambda_S}\vec{Y}^\sigma_{\geq\lambda_T}\right) \notag\\
    &= \E_{\sigma,\lambda_T} \left[\frac{2}{m}\sum_{i=m/2+1}^{m} \I_Q\left(X^\sigma_i;M\mid \vec{X}^\sigma_{<i}\vec{Y}^\sigma_{\geq \lambda_T}\right) \right] \notag\\
    &= \frac{2}{m} \E_{\sigma,\lambda_T} \I_Q\left(\vec{X}^\sigma_{>m/2};M\mid \vec{X}^\sigma_{\leq m/2}\vec{Y}^\sigma_{\geq \lambda_T}\right). \label{eq:expected-FC}
\end{align}
For fixed $\sigma$ and $\lambda_T$, let $\vec{A}=\vec{X}^\sigma_{>m/2}$, $\vec{J}=\vec{X}^\sigma_{\leq m/2}\vec{Y}^\sigma_{\geq\lambda_T}$, $\vec{B}=\vec{Y}^\sigma_{<\lambda_T}$. Note that for any assignment of positive probability, 
\begin{equation*}
    P\left(\vec{Y}^\sigma_{<\lambda_T} = \vec{b} \mid \vec{X}^\sigma=\vec{x}^\sigma, \vec{Y}^\sigma_{\geq \lambda_T} = \vec{y}^\sigma_{\geq \lambda_T}, M=\tau\right) 
    = \prod_{j<\lambda_T} \mu(Y^\sigma_j =b_j \mid X^\sigma_j =x^\sigma_j).
\end{equation*}
Moreover, as $\lambda_T\leq m/2$, $X^\sigma_{<\lambda_T}$ is revealed in $\vec{J}$, hence $\vec{B}$ is conditionally independent of $(\vec{A},M)$ given $\vec{J}$.

By mutual information chain rule,
\[
    \I_P(\vec{A};\vec{B}M\mid \vec{J}) = \I_P(\vec{A};M\mid \vec{J}) + \I_P(\vec{A};\vec{B}\mid M\vec{J}) = \I_P(\vec{A};\vec{B}\mid \vec{J}) + \I_P(\vec{A};M\mid \vec{B}\vec{J}).
\]
The conditional independence conditions of $\vec{B}$ gives
\[
    \I_P(\vec{A};\vec{B}\mid \vec{J}) = \I_P(\vec{A};\vec{B}\mid M\vec{J}) = 0,
\]
hence $\I_P(\vec{A};M\mid \vec{J}) = \I_P(\vec{A};M\mid \vec{B}\vec{J})$, or equivalently 
\begin{equation}\label{eq:FC1}
    \I_P(\vec{X}^\sigma_{>m/2}; M\mid \vec{X}^\sigma_{\leq m/2}\vec{Y}^\sigma_{\geq \lambda_T}) 
    = \I_P(\vec{X}^\sigma_{>m/2}; M\mid \vec{X}^\sigma_{\leq m/2}\vec{Y}).
\end{equation}
Applying \cref{eq:FC1}, non-negativity of mutual information, and \cref{eq:MI-bound} in order,
\begin{equation}\label{eq:FC2}    
    \I_P(\vec{X}^\sigma_{>m/2}; M\mid \vec{X}^\sigma_{\leq m/2}\vec{Y}^\sigma_{\geq \lambda_T}) 
    = \I_P(\vec{X}^\sigma_{>m/2}; M\mid \vec{X}^\sigma_{\leq m/2}\vec{Y})
    \leq \I_P(\vec{X};M\mid \vec{Y})
    \leq \gamma.
\end{equation}
To upper bound $\E_c[\FC_c]$, it remains to deduce a bound with $Q=P(\cdot\mid \scW)$ in place of $P$. First, by considering $\1_{\scW}$ and non-negativity,
\begin{align}
    \I_P(\vec{A};M\mid \vec{J}\1_{\scW}) 
    &= P(\scW) \I_P(\vec{A};M\mid \vec{J},\1_{\scW}=1)+(1-P(\scW)) \I_P(\vec{A};M\mid \vec{J},\1_{\scW}=0) \notag\\
    &\geq P(\scW) \I_Q(\vec{A};M\mid \vec{J}) \label{eq:FC3}.
\end{align}
Next, we evaluate $\I_P(\vec{A};M\mid \vec{J}\1_{\scW})$ by chain rules in two ways:
\[
    \I_P(\vec{A};M\1_{\scW}\mid \vec{J}) 
    = \I_P(\vec{A};M\mid \vec{J}) + \I_P(\vec{A};\1_{\scW}\mid M\vec{J}) 
    = \I_P(\vec{A};\1_{\scW}\mid \vec{J}) + \I_P(\vec{A};M\mid \vec{J}\1_{\scW}).
\]
We obtain
\begin{align*}
    \I_P(\vec{A};M\mid \vec{J}\1_{\scW}) 
    &= \I_P(\vec{A};M\mid \vec{J}) + \I_P(\vec{A};\1_{\scW}\mid M\vec{J}) - \I_P(\vec{A};\1_{\scW}\mid \vec{J})\\
    &\leq \I_P(\vec{A};M\mid \vec{J}) + \H(\1_{\scW}\mid M\vec{J})\\
    &\leq \I_P(\vec{A};M\mid \vec{J}) + \ent(P(\scW))
\end{align*}
Combining with \cref{eq:expected-FC,eq:FC2,eq:FC3} give
\begin{equation}\label{eq:FC_bound}
    \E_c[\FC_c]= \frac{2}{m}\E_{\sigma,\lambda_T} \I_Q(\vec{A};M\mid \vec{J})\leq \frac{2(\gamma+\ent(P(\scW)))}{mP(\scW)}\leq \frac{2(\gamma+\ent(p))}{mp}.
\end{equation}
The last inequality uses the fact that $u\mapsto (\gamma+\ent(u))/u$ is non-increasing on $(0,1]$.

\paragraph*{Backward view.}
By a similar reasoning,
\begin{equation}\label{eq:BC1}
    \E_c[\BC_c] = \E_{\sigma,\lambda_S,\lambda_T} \I_Q\left(Y^\sigma_{\lambda_T};M\mid \vec{X}^\sigma_{\leq \lambda_S}\vec{Y}^\sigma_{> \lambda_T}\right) = \frac{2}{m}\E_{\sigma,\lambda_S} \I_Q\left(\vec{Y}^\sigma_{\leq m/2};M\mid \vec{X}^\sigma_{\leq\lambda_S}\vec{Y}^\sigma_{> m/2}\right).
\end{equation}
For fixed $\sigma$ and $\lambda_S$, let
\(
    \vec A=\vec X^\sigma_{>\lambda_S},
    \vec J=\vec X^\sigma_{\le\lambda_S}
            \vec Y^\sigma_{>m/2},
    \vec B=\vec Y^\sigma_{\le m/2}.
\)
By the definition of $P$, for any assignment of positive probability
\[
P(
    \vec Y^\sigma_{\le m/2}=\vec b
    \mid
    \vec X^\sigma=\vec x^\sigma,
    \vec Y^\sigma_{>m/2}=\vec y^\sigma_{>m/2},
    M=\tau
)=
\prod_{j\le m/2}
\mu(Y^\sigma_j=b_j\mid X^\sigma_j=x^\sigma_j).
\]
Since $\lambda_S>m/2$, $\vec X^\sigma_{\leq m/2}$ is revealed in $\vec{J}$, hence $\vec{B}$ is conditionally independent of $(\vec{A},M)$ given $\vec{J}$ under $P$. 
In particular,
\[
    \I_P\left(
        \vec Y^\sigma_{\le m/2};M
        \mid 
        \vec X^\sigma_{\le\lambda_S}
        \vec Y^\sigma_{>m/2}
    \right)=0.
\]
Applying \cref{lem:cond-ind-MI-bound} on \cref{eq:BC1} gives
\[
    \E_c[\BC_c] = \frac{2}{m}\E_{\sigma,\lambda_S} \I_Q\left(\vec{Y}^\sigma_{\leq m/2};M\mid \vec{X}^\sigma_{\leq\lambda_S}\vec{Y}^\sigma_{> m/2}\right) \leq \frac{2\L(P(\scW))}{m}\leq \frac{2\L(p)}{m}.
\]

\paragraph*{Error bound.}
The expected error bound can be bounded similarly. 
By data processing inequality for mutual information and chain rule
\begin{align*}
    \E_c[\RD_c] 
    &=\E_c \I_Q(X_i;Z_i\mid MY_i,(\vec{X}_S,\vec{Y}_T)=\vec{r})\\
    &\leq\E_c \I_Q(X_i;\vec{Z}\mid MY_i,(\vec{X}_S,\vec{Y}_T)=\vec{r})\\
    &=\E_{\sigma,\lambda_S,\lambda_T} \I_Q(X^\sigma_{\lambda_S};\vec{Z}\mid \vec{X}^\sigma_{<\lambda_S}Y^\sigma_{\geq \lambda_T}M)\\
    &=\frac{2}{m}\E_{\sigma,\lambda_T} \I_Q(\vec{X}^\sigma_{>m/2};\vec{Z}\mid \vec{X}^\sigma_{\leq m/2}Y^\sigma_{\geq \lambda_T}M).
\end{align*}
As in the above arguments, $\vec{X}^\sigma_{>m/2}$ and $\vec{Z}$ are conditionally independent given $\vec{X}^\sigma_{\leq m/2}Y^\sigma_{\geq \lambda_T}M$, so \cref{lem:cond-ind-MI-bound} yields
\[
    \E_c[\RD_c] \leq \frac{2\L(P(\scW))}{m}\leq \frac{2\L(p)}{m}.
\]

\subsection{Input Distribution Guarantees}
We now handle the expected closeness of the new input distribution for $f$. Fix $(i,S,T)$, note that under $P$ the pair $X_iY_i$ is distributed identically to $\mu$ and is independent to $\Vec{X}_S\Vec{Y}_T$. 
By the definition of $\mu'_c$ and independence under $P$,
\[
    \KL(\mu'_c\Vert \mu) = \KL\left(\doublefrac{Q[X_iY_i\mid (\Vec{X}_S,\Vec{Y}_T)=\vec{r}]}{P[X_iY_i\mid (\Vec{X}_S,\Vec{Y}_T)=\vec{r}]}\right)
\]
The chain rule on $\KL(Q[X_iY_i\Vec{X}_S\Vec{Y}_T]\Vert P[X_iY_i\Vec{X}_S\Vec{Y}_T])$ yields
\[
    \E_{\vec{r}\sim Q[\Vec{X}_S\Vec{Y}_T]} \KL(\mu'_c\Vert \mu) = \KL(Q[X_iY_i\Vec{X}_S\Vec{Y}_T]\Vert P[X_iY_i\Vec{X}_S\Vec{Y}_T]) - \KL(Q[\Vec{X}_S\Vec{Y}_T]\Vert P[\Vec{X}_S\Vec{Y}_T]).
\]
By data processing inequality, we have
\[
    \E_{\vec{r}\sim Q[\Vec{X}_S\Vec{Y}_T]} \KL(\mu'_c\Vert \mu)\leq \KL(Q\Vert P) = \L(P(\scW))\leq \L(p)
\]
and thus $\E_c \KL(\mu'_c\Vert \mu) \leq \L(p)$.

To establish the expected KL conditional closeness, we bound the conditional divergence on $Y\mid X$ via the backward view, and bound the conditional divergence on $X\mid Y$ via the forward view.

From the backward view, for a fixed configuration $c=(i,S,T,\vec{r})$, $\mu'_c = Q[X_{\sigma(\lambda_T)}Y_{\sigma(\lambda_T)}\mid (\vec{X}_S,\vec{Y}_T) = \vec{r}]$. By the definition of conditional KL divergence,
\begin{align}
    &\CRE{\mu'_c[Y\mid X]}{\mu[Y\mid X]}{\mu'_c[X]} \notag\\
    &= \E_{x\sim Q[X_{\sigma(\lambda_T)}\mid (\vec{X}_S,\vec{Y}_T) = \vec{r}] } \KL\left(\doublefrac{ Q[Y_{\sigma(\lambda_T)}\mid X_{\sigma(\lambda_T)}=x, (\vec{X}_S,\vec{Y}_T) = \vec{r}] }{ \mu[Y\mid X=x] }\right)\notag\\
    &= \E_{x\sim Q[X_{\sigma(\lambda_T)}\mid (\vec{X}_S,\vec{Y}_T) = \vec{r}] } \KL\left(\doublefrac{ Q[Y_{\sigma(\lambda_T)}\mid \vec{X}^\sigma_{\leq\lambda_S}\vec{Y}^\sigma_{>\lambda_T} ] }{ \mu[Y\mid X=x] }\right)\label{eq:cond-Y|X},
\end{align}
where the last equality follows from the observation that $X_{\sigma(\lambda_T)}=x, (\vec{X}_S,\vec{Y}_T) = \vec{r}$ fix the values of $\vec{X}^\sigma_{\leq\lambda_S}\vec{Y}^\sigma_{>\lambda_T}$. Writing $\vec{J}_{\lambda_S}(\lambda_T) = \vec{X}^\sigma_{\leq\lambda_S}\vec{Y}^\sigma_{>\lambda_T}$, by the definition of $P$, we also have $P[Y_{\sigma(\lambda_T)}\mid \vec{J}_{\lambda_S}(\lambda_T)] = \mu[Y\mid X_{\sigma(\lambda_T)}]$. 

Averaging \cref{eq:cond-Y|X} over $\vec{r}\sim Q[\Vec{X}_S\Vec{Y}_T]$ yields the equality
\[
    \E_{\vec{r}\sim Q[\Vec{X}_S\Vec{Y}_T]}
    \CRE{
        \mu'_c[Y\mid X]
    }{
        \mu[Y\mid X]
    }{
        \mu'_c[X]
    }
    =
    \CRE{
        Q[Y_{\sigma(\lambda_T)}\mid\vec{J}_{\lambda_S}(\lambda_T)]
    }{
        P[Y_{\sigma(\lambda_T)}\mid\vec{J}_{\lambda_S}(\lambda_T)]
    }{
        Q[\vec{J}_{\lambda_S}(\lambda_T)]
    }.
\]
Note that since $\lambda_S > m/2\geq \lambda_T$, we have $\vec{J}_{\lambda_S}(\lambda_T) = \vec{J}^* (Y_{\sigma(\lambda_T+1)}\ldots Y_{\sigma(m/2)})$, where $\vec{J}^*=\vec{X}^\sigma_{\leq\lambda_S} \vec{Y}^\sigma_{> m/2}$. By chain rule,
\[
    \CRE{
        Q[\vec{Y}^\sigma_{\leq m/2}\mid\vec{J}^*]
    }{
        P[\vec{Y}^\sigma_{\leq m/2}\mid\vec{J}^*]
    }{
        Q[\vec{J}^*]
    }
    =
    \sum_{\lambda_T=1}^{m/2}
    \CRE{
        Q[Y_{\sigma(\lambda_T)}\mid\vec{J}_{\lambda_S}(\lambda_T)]
    }{
        P[Y_{\sigma(\lambda_T)}\mid\vec{J}_{\lambda_S}(\lambda_T)]
    }{
        Q[\vec{J}_{\lambda_S}(\lambda_T)]
    }.
\]
hence
\[
    \E_c \CRE{\mu'_c[Y\mid X]}{\mu[Y\mid X]}{\mu'_c[X]} = \frac{2}{m} \E_{\sigma,\lambda_S} \CRE{Q[\vec{Y}^\sigma_{\leq m/2}\mid\vec{J}^*]}{ P[\vec{Y}^\sigma_{\leq m/2}\mid\vec{J}^* ] }{ Q[ \vec{J}^* ] }
\]
Lastly, by chain rule, non-negativity and data processing inequality,
\begin{align*}
    \CRE{Q[\vec{Y}^\sigma_{\leq m/2}\mid\vec{J}^*]}{ P[\vec{Y}^\sigma_{\leq m/2}\mid\vec{J}^* ] }{ Q[ \vec{J}^* ] } &= \KL\left(\doublefrac{Q[\vec{Y}^\sigma_{\leq m/2}\vec{J}^*]}{ P[\vec{Y}^\sigma_{\leq m/2}\vec{J}^* ] }\right) - \KL(Q[\vec{J}^*]\Vert P[\vec{J}^*]) \\
    &\leq \KL(Q\Vert P)-0\leq \L(P(\scW))\leq \L(p).
\end{align*}
This shows that $\mu'_c$ is $O(\L(p)/m)$-conditionally-close to $\mu$ on $Y\mid X$ in expectation over $c$. The conditional closeness on $X\mid Y$ can be shown by a symmetric argument.

To conclude the proof, we have shown expectation upper bounds of the six quantities: $\FC_c$, $\BC_c$, lossy protocol error bound $e_c$, $\KL(\mu_c',\mu)$, and two conditional KL divergence bounds on $(X,Y)$.
By Markov's inequality and a union bound, all these six quantities are bounded above by 7 times their respective expectations with probability at least $1/7$. For such a configuration $c^*$, the input distribution $\mu'_{c^*}$ and the lossy protocol $\Pi_{c^*}$ satisfy all the required properties in \cref{lem:direct-product}. This concludes the proof of the lemma.
\section{Forward Information Cost Lower Bound of $\UR_{U;\alpha}$}\label{app:Lemma5}
In this section, we present the proof of \cref{lem:LB-UR}. As mentioned previously, the proof is adapted from \cite[Lemma 4.5]{nelson_yu:SODA.2019} while untangling the dependencies between parameters in the original proof. We first recall the lemma statement in our notation.

\LBUR*

We first prove a lower bound on the forward information cost of a lossy protocol for an auxiliary instance $\UR_{V;\beta}$, where
\[
s_0\le s\le \beta/2,\quad 
V\ge 10\beta,\quad
\epsilon+\eta+\theta\le 2^{-s}.
\]
The hard distribution $\nu$ for $\UR_{V;\beta}$ follows the construction in \cite{nelson_yu:SODA.2019}, with the universe size, Alice’s input size, and $\L(\delta)$ replaced by $V$, $\beta$, and $s$, respectively. Set
\[
p=\frac{20}{s},\quad
R=\floorB{\frac{s}{400}\log\frac{20\beta}{s}},\quad
r_i=\floorB{\beta[1-(1-p)^i]}
\quad\text{for }0\le i<R,\qquad r_R=\beta.
\]
Under $\nu$, Alice’s input $A'$ is sampled uniformly from $\binom{[V]}{\beta}$, $J$ is sampled independently and uniformly from $\{0,\ldots,R-1\}$, and Bob’s input $B'$ is sampled uniformly from $\binom{A'}{r_J}$. For sufficiently large $s_0$, these choices guarantee that $r_{i+1}-r_i\ge1$ for every $0\le i<R$.

We adapt the encoding argument in \cite[Appendix B]{nelson_yu:SODA.2019} and refer there for the complete encoding and decoding procedures. Here we give a high-level description of the features used in our analysis. The lower bound by Nelson and Yu relies on three estimates. Claim 1 of \cite{nelson_yu:SODA.2019} bounds the mutual information between Bob's initial set and Alice's set. Claim 2 of \cite{nelson_yu:SODA.2019} shows that the decoder succeeds with high probability on a fresh query using the original message. Claim 3 of \cite{nelson_yu:SODA.2019} bounds how much adaptivity can increase the probability of any particular query set relative to uniform sampling at the same size. Our analysis follows the same approach, and for clarity we present the essential part of the argument in our notation.

\paragraph*{Encoding Procedure.}
Let $\nu'$ be $(\zeta,\eta)$-KL-close to $\nu$. Sample $(A',B')\sim \nu'$ and $M\sim \Msg(\cdot \mid A',B')$, and then independently sample $\tdB\sim \nu(\cdot \mid A')$ given $A'$. The set to be encoded $A$ is the realized set of the random variable $A'$. Since $\KL(\nu'\Vert\nu)\leq \zeta<\infty$, this ensures that $A$ always lies in the support of $\nu[A']$, hence $\abs{\tdB}=r_{i'}$ for some $i'\in \set{0,\ldots,R-1}$.

Initialize the successful recovery set $C^{(i')} = \emptyset$ and the decoder query set $Q^{(i')}=\tdB$. In each round $i=i',\ldots,R-1$, the encoder reuses $M$ to sample $z_i\sim \Dec(\cdot \mid MQ^{(i)})$, where $\Dec$ is the decoder from the lossy protocol $\tilde{\Pi}$. It first sets $C^{(i+1)}=C^{(i)}$ and $Q^{(i+1)}=Q^{(i)}$, adds $z_i$ to $C^{(i+1)}$ and $Q^{(i+1)}$ if it is in $A\setminus Q^{(i)}$, and then fills $Q^{(i+1)}$ to size $r_{i+1}$ taking the remaining elements of $A$ in the order given by a publicly sampled uniformly random permutation. By reusing $M$, the encoder avoids explicitly encoding the elements of $C^{(R)}$. We therefore seek a lower bound on the number of successful rounds $N\defeq \abs{C^{(R)}}$.

\paragraph*{Analysis of the Auxiliary Instance.}
Denote $\rho$ and $\tilde{\rho}$ the marginals on $A'B'M$ and $A'\tdB M$ respectively, that is
\begin{align*}
    \rho(A',B',M) &= \nu'(A',B') \Msg(M\mid A',B'),\\
    \tilde{\rho}(A',\tdB,M) &= \nu'(A')\nu(\tdB\mid A')\Msg(M\mid A').
\end{align*}
In particular, $\rho[A'M] = \tilde{\rho}[A'M]$.

Using the fact that $\tilde{B}$ and $M$ are conditionally independent given $A'$ under $\tilde{\rho}$, a KL-divergence chain rule derivation similar to that in \cref{eq:cond-ind} gives
\begin{equation}\label{eq:KL_rho}
    \KL(\rho\Vert \tilde{\rho}) = \I_\rho(B';M\mid A') +\CRE{\nu'[B'\mid A']}{\nu[B'\mid A']}{\nu'[A']} \leq \theta+\eta.
\end{equation}
Denote $e_\rho$ and $e_{\tilde{\rho}}$ the protocol error probability under $\rho$ and $\tilde{\rho}$ respectively. By applying data processing inequality on \cref{eq:KL_rho}, we have $\KL(\text{Bern}(e_\rho)\Vert \text{Bern}(e_{\tilde{\rho}}))\leq \theta+\eta$.
By log-sum inequality (see e.g. \cite[Theorem 2.7.1]{CT05}),
\[
    \KL(\text{Bern}(e_\rho)\Vert \text{Bern}(e_{\tilde{\rho}})) +e_{\rho} = e_\rho \log \frac{e_\rho}{e_{\tilde{\rho}}/2} + (1-e_\rho) \log \frac{1-e_\rho}{1-e_{\tilde{\rho}}} \geq \log \frac{1}{1-e_{\tilde{\rho}}/2},
\]
by the inequality $\ln(1+x)\leq x$ for $x\in (-1,\infty)$ we obtain
\[
    e_{\tilde{\rho}} \leq 2\ln 2 \cdot \log \frac{1}{1-e_{\tilde{\rho}}/2} \leq 2\ln 2(\epsilon+\theta+\eta)\leq 2^{-s+1}.
\]
This recovers the result in Claim 2 of \cite{nelson_yu:SODA.2019}.

For fixed $A'=A$, $M=m$ and starting level $I=i'$, by exactly the same calculations in Claim 3 of \cite{nelson_yu:SODA.2019}, for any $i\in [i',R]$ and $Q\in \binom{A}{r_i}$,
\begin{equation}\label{eq:almost-uniformity}
    \Pr[Q^{(i)}=Q\mid A,m,i'] \leq \frac{1}{\binom{\beta}{r_i}} \cdot 2^{0.9s}.
\end{equation}
Denote $e_i(A,m)$ to be the error probability of the decoder on a uniform query of size $r_i$:
\[
    e_i(A,m) \defeq \frac{1}{\binom{\beta}{r_i}} \sum_{Q\in \binom{A}{r_i}} \Pr_{z\sim \Dec(\cdot\mid mQ)}[z\notin A\setminus Q].
\]
Since the starting level of the resampled initial query is uniform, 
\[
    e_{\tilde{\rho}} = \E_{\tilde{\rho}} \left[\frac{1}{R} \sum_{i=0}^{R-1} e_i(A',M)\right],
\]
and the expected number of rounds is $(R+1)/2$.
Combining with \cref{eq:almost-uniformity} yields
\[
    \E_{\tilde{\rho}}[N] \geq \frac{R+1}{2} - 2^{0.9 s} \E_{\tilde{\rho}} \left[\sum_{i=0}^{R-1} e_i(A',M)\right] \geq \frac{R}{2} - 2^{0.9 s} R e_{\tilde{\rho}} \geq \left(\dfrac{1}{2} - 2\cdot 2^{-0.1s}\right) R .
\]
Note that the encoder uses the same procedure and randomness distribution once the triple $A'=A$, $B'=B$, $M=m$ is fixed. Therefore
\begin{align*}
    \E_{\tilde{\rho}} [N] - \E_{\rho} [N]
    &= \sum_{A,B,m} \E[N\mid A,B,m] (\tilde{\rho}(A,B,m) - \rho(A,B,m))\\
    &\leq R \sum_{A,B,m} \max\set{\tilde{\rho}(A,B,m) - \rho(A,B,m), 0}\\
    &=R\cdot  \TV(\rho,\tilde{\rho})
\end{align*}
Applying Pinsker's inequality and \cref{eq:KL_rho},
\begin{align*}
    \E_{\rho}[N] &\geq \E_{\tilde{\rho}}[N] - R \cdot \TV(\rho,\tilde{\rho}) \\
    &\geq \left(\dfrac{1}{2} - 2\cdot 2^{-0.1s}\right) R -  R \sqrt{\frac{\ln 2}{2}(\theta+\eta)} \\
    &\geq \left(\dfrac{1}{2} - 2\cdot 2^{-0.1s}-\frac{3}{5} \cdot 2^{-0.5s}\right) R,
\end{align*}
which is at least $0.49 R$ for $s$ above some sufficiently large constant $s_0$.

To complete the proof for the auxiliary instance, Claim 1 and the subsequent calculations in \cite{nelson_yu:SODA.2019} bound the forward information cost in two ways. Under $\nu$, conditioning on $B'=B$, $A'$ is uniform over all $\beta$-element supersets of $B$. Together with conditional closeness between $\nu$ and $\nu'$, it follows that
\[
    \H_{\nu'}(A'\mid B')\geq  \E_{\nu'} \left[\log \binom{V-\abs{B'}}{\beta-\abs{B'}}\right] - \eta.
\]
Using this revised bound on $\H_{\nu'}(A'\mid B')$, exactly the same binomial estimate following Claim 1 of \cite{nelson_yu:SODA.2019} implies that
\[
    \I_\rho (A'; M\mid B) \geq \E_\rho [N] \log \frac{V-\beta}{\beta} - R - \eta =\Omega\left(R \log \frac{V}{\beta}\right) = \Omega\left(s\log \frac{\beta}{s} \log \frac{V}{\beta}\right).
\]
Here, the second-to-last bound uses $\E_\rho [N]\geq 0.49 R$, $V\geq 10\beta$ and $\eta\leq 2^{-s}$, and the last bound uses $R=\Theta(s\log (\beta/s))$.

\paragraph*{General bound.}
For the general case with parameters $U$, $\alpha$ and $h$, a padding argument yields the required bound. Concretely, we set $s=h/1000$, $\beta=\floor{\sqrt{\alpha h}/100}$ and $V=U-\alpha+\beta$. Let $P$ be the last $\alpha-\beta$ elements in $[U]$, each auxiliary input $(A',B')$ of $\UR_{V;\beta}$ is mapped to $(A,B)=(A'\cup P,B'\cup P)$. The new instance satisfies $\abs{A}=\alpha$, $B\subsetneq A$ and $A\setminus B = A'\setminus B'$. Using the lossy protocol forward information bound for $\UR_{V;\beta}$, we obtain
\[
    \I(A;M\mid B) = \I(A';M\mid B') =\Omega\left(s\log \frac{\beta}{s} \log \frac{V}{\beta} \right) =\Omega\left(h\log \frac{\alpha}{h} \log \frac{U}{h} \right).
\]

\section{Application to Worst-Case Dynamic Connectivity}
\label{app:dynamic}

We prove the dynamic spanning forest statement in \Cref{thm:dynamic}
using the nested-forest framework of
Kapron, King, and Mountjoy~\cite{KKM13} and
Wang~\cite{Wang15}, with cutset structures as in
Gibb et al.~\cite{GKKT15}.

The analysis rests on a deletion-sensitive bound: in the ideal oracle
process, at most $2k$ bottom-layer trees have outgoing edges in the
current graph (\Cref{lem:appd-bottom}). This allows us to use
$D=\Theta(\log(1+k))$ recovery layers, with
$q=\Theta(1+\L(\delta)/\log(1+k))$ independent cutset copies per layer.
The copies on each layer maintain the same forest and use independent
randomness to amplify recovery success. The total number of cutset
structures is therefore
\[
 L:=Dq=\Theta\bigl(\log(1+k)+\L(\delta)\bigr).
\]

\subparagraph*{Notation.}
Throughout this appendix, $D$ denotes the number of recovery layers
(unrelated to the deletion multiset of \Cref{sec:kec-deter}), and $b$
denotes both the number of sampling levels in a cutset copy and its
hash width.
The parameter $\delta$ denotes the failure probability, while
$\delta_G(C)$ denotes the cut induced by a vertex set $C$ in $G$.

\begin{theorem}[\Cref{thm:dynamic}, restated]
\label{thm:appd-dynamic}
Fix a constant $\kappa>0$, and let $n\ge2$, $1\le k\le n^2$, and
$n^{-\kappa}\le\delta<1/2$. There is a randomized data structure on a
word RAM with $\Theta(\log n)$-bit words that, given $k$ and $\delta$,
maintains a forest while processing valid edge insertions and
deletions in an initially empty simple undirected graph on $[n]$.
For every oblivious update sequence containing at most $k$ deletions,
with probability at least $1-\delta$, the maintained forest is a
spanning forest of the current graph after every update.
Each insertion takes $O_\kappa(L\log^2 n)$ worst-case time, and each
deletion takes $O_\kappa(L\log^2 n\log(1+k))$ worst-case time, where
\[
 L=\Theta\bigl(\log(1+k)+\L(\delta)\bigr).
\]
Connectivity queries take $O(\log n/\log\log n)$ worst-case time.
\end{theorem}

\begin{remark}[Model and scope]
\label{rem:appd-model}
An insertion is valid if the edge is absent, and a deletion is valid
if the edge is present. The deletion budget counts deletion
operations, including repeated deletions of an edge that has been
reinserted.

All randomness is sampled during preprocessing and reused throughout
the update sequence. The update and query time bounds hold for every
outcome of these random choices. Preprocessing selects the prime
of~\eqref{eq:appd-prime} in expected
$\operatorname{poly}(\log n)$ time
(\Cref{sec:appd-complexity}) and initializes the
structures in $O(nL\log n)$ time; these costs are not charged to updates.

The restriction $k\le n^2$ is used only to bound the field size.
Together with $\delta\ge n^{-\kappa}$, it ensures that the prime
of~\eqref{eq:appd-prime} satisfies
$\log p=O((1+\kappa)\log n)$.
Thus a field element occupies $O(1+\kappa)$ words, and each field
operation takes $O((1+\kappa)^2)$ word operations. This is the factor
hidden by the $O_\kappa(\cdot)$ notation.
\end{remark}

We first describe the recovery primitive and the update rules.
We then adapt Wang's oracle-coupling argument to prove simultaneous
correctness, accounting separately for recovery failures and
fingerprint errors, and establish the worst-case running time and
space bounds.

\subsection{The recovery primitive and amplification}
\label{sec:appd-cutset}

For a graph $G$ and a vertex set $C$, write
\[
 \delta_G(C)=\{uv\in E(G):|\{u,v\}\cap C|=1\}.
\]
A cutset structure maintains a forest $F$ together with
graph-incidence sketches. It supports graph-edge insertions and
deletions, forest-edge links and cuts, and recovery of an edge
leaving a tree of $F$.

Each copy uses a fixed sampling seed to select sketch coordinates
and a fixed fingerprint point to verify recovery candidates.
Recovery is a deterministic function of the current graph, the
queried vertex set, and these random choices; recovery calls neither
consume fresh randomness nor modify the sketches.
For the analysis, we first define an exactly sound ideal recovery
rule that uses only the sampling seed. Fingerprint errors in its
implementation are analyzed separately in
\Cref{sec:appd-soundness}.

Let $N=\binom n2$, and use the edge identifiers
$\operatorname{id}:\binom{[n]}2\to[N]$ and signed incidence vectors
$x^{(v)}$ from \Cref{subsec:rand-upper-recovery}.
Identifying coordinates with edges via $\operatorname{id}$, we have
\[
 x(C):=\sum_{v\in C}x^{(v)},
 \qquad
 \operatorname{supp}(x(C))=\delta_G(C).
\]

\subparagraph*{Sampling levels.}
We retain the full range of
\[
 b:=\lceil\log_2(2N)\rceil
\]
sampling levels, rather than truncating at the deletion budget,
and also use $b$ as the hash width.
All copies use a common representation of $\mathbb F_{2^b}$
by $b$-bit integers, defined by a deterministically chosen
irreducible polynomial
(\Cref{sec:appd-complexity}).
Since $N<2^b$, we identify $[N]$ with a subset of this field.

For each copy, choose independent uniform
$\beta,\beta'\in\mathbb F_{2^b}$ and define
\[
 h(j)=\beta j+\beta'.
\]
This affine hash family is pairwise independent
\cite{carter_wegman:JCSS.1979}.
Interpreting hash values as integers in $\{0,\ldots,2^b-1\}$,
define
\[
 S_\ell:=\{e:h(\operatorname{id}(e))<2^{b-\ell}\},
 \qquad \ell=1,\ldots,b.
\]
Thus, for every $\ell\in[b]$, each edge belongs to $S_\ell$
with probability exactly $2^{-\ell}$, and the sampling events
for distinct edges are pairwise independent.
The choice of hash width ensures that these probabilities are
exact at all $b$ levels.

\subparagraph*{Ideal recovery and amplification.}
The ideal decoder examines the levels in increasing order.
It returns the unique edge of $\delta_G(C)\cap S_\ell$
at the first level where this intersection is a singleton,
and returns $\bot$ if no such level exists.
For every sampling seed, every returned edge therefore belongs
to $\delta_G(C)$; in particular, an empty cut always yields $\bot$.

Fix a graph $G$ and a vertex set $C$ with
$s:=|\delta_G(C)|\ge1$.
The level
\[
 \ell^*=\lceil\log_2(2s)\rceil\le b
\]
has sampling probability $\rho:=2^{-\ell^*}$ satisfying
\[
 \frac{1}{4s}<\rho\le\frac{1}{2s}.
\]
By pairwise independence and a union bound, a specified cut edge
is the unique sampled cut edge with probability at least
$\rho-(s-1)\rho^2$.
Summing over the disjoint singleton events gives
\[
 \Pr\bigl[|\delta_G(C)\cap S_{\ell^*}|=1\bigr]
 \ge s\rho-s(s-1)\rho^2
 \ge \frac{s\rho}{2}
 >\frac18.
\]
Hence a single copy succeeds with probability greater than $1/8$.

At each recovery layer, use $q$ independent copies and return
the first non-$\bot$ answer in a fixed copy order.
For every fixed nonempty cut, amplified ideal recovery fails
with probability at most
\begin{equation}
 \varepsilon=(7/8)^q.
 \label{eq:appd-epsilon}
\end{equation}
Sampling seeds are mutually independent across copies and layers.
Trees on the same layer share the same copies, so their recovery
outcomes may be dependent.
The fixed-cut guarantee above applies whenever the cut is
independent of the sampling seeds used for recovery;
its use for trees generated by the algorithm is justified in
\Cref{sec:appd-oracle}.

\subparagraph*{Implementation and operation costs.}
The implementation uses the two measurements and singleton
fingerprints of \Cref{subsec:rand-upper-recovery}.
Each copy has one fingerprint point $\alpha\in\mathbb F_p$,
shared across its sampling levels.
Conditional on the prime $p$ of~\eqref{eq:appd-prime},
these points are independent and uniform, and independent
of all sampling seeds.
Copies and levels are tested in the same fixed order as in
ideal recovery.

Each copy stores $O(b)$ field elements per vertex.
A balanced Euler-tour representation maintains their sum over
each forest tree, together with the tree size.
A graph update evaluates $h$ and the fingerprint coefficient
once, then updates at most $b$ sampled coordinates along
$O(\log n)$ aggregate nodes.
The hash evaluation takes $O(b)$ word operations using
shift-and-xor multiplication and reduction, and the fingerprint
coefficient is computed by repeated squaring.
A forest link or cut changes $O(\log n)$ aggregate nodes,
each storing $O(b)$ field elements.
Recovery examines at most $b$ candidates, each requiring
$O_\kappa(\log n)$ time for exponentiation and
edge-identifier decoding.

Since $b=O(\log n)$, each graph update, forest link or cut,
and recovery takes $O_\kappa(\log^2 n)$ worst-case time
per copy. An amplified cutset operation therefore takes
$O_\kappa(q\log^2 n)$ worst-case time.

\subsection{The nested forests and update rules}
\label{sec:appd-algorithm}

Maintain nested forests
\[
 F_0\subseteq F_1\subseteq\cdots\subseteq F_D
\]
and $q$ recovery copies for each $F_i$, $0\le i<D$.
The reported forest is $F_D$. A tree of $F_i$ is \emph{active} if it
has an outgoing edge in the current graph. It is \emph{merged} if its
vertex set is strictly contained in a tree of $F_{i+1}$; otherwise it
is \emph{unmerged}, that is, its vertex set is also a tree of
$F_{i+1}$. Define an edge's \emph{tier} as the least layer containing
it. Initially every forest has no edges.

\subparagraph*{Insertion of $e=uv$.}
Update the graph-incidence sketches in every recovery copy. If $u$
and $v$ are in different trees of $F_D$, link $e$ into all forests
$F_0,\ldots,F_D$, with tier $0$. Otherwise change no forest edge.
No recovery is performed on an insertion.

\subparagraph*{Deletion of $e=uv$.}
Remove $e$ from the graph-incidence sketches and from every forest
containing it. Process layers $i=0,\ldots,D-1$ in increasing order.
At layer $i$, process $u$ and then $v$. For the current $F_i$ tree
$C$ containing the processed endpoint, skip recovery if $C$ is
already merged at layer $i+1$; this is decided by comparing the sizes
of the trees of $F_i$ and $F_{i+1}$ containing the endpoint.
Otherwise apply amplified recovery to $C$. If it returns an edge
$e'=u'v'$ with exactly one endpoint in $C$, insert $e'$ into every
forest strictly above layer $i$, using the cycle-exchange rule below.
Invalid edge identifiers and candidates that do not leave $C$ are
rejected.

\subparagraph*{Cycle exchange.}
Let $r>i$ be the first layer connecting $u'$ and $v'$, if such a layer
exists. Among the edges of tier $r$ on their unique path in $F_r$,
let $f$ be the one with least edge identifier. Remove $f$ from all
forests $F_r,\ldots,F_D$, and then insert $e'$ into all forests
$F_{i+1},\ldots,F_D$, with tier $i+1$. If no such $r$ exists, simply
perform the insertions. Since every edge of the $F_r$ path has tier
at most $r$, the edge $f$ is a maximum-tier edge on the path. It need
not directly join the two endpoint components of $F_{r-1}$.

The maximum-tier rule is the path-exchange interpretation of Wang's
framework. All forest-edge changes are applied to every recovery copy
of the affected layer. In particular, an edge recovered at layer $i$
is linked into each higher layer's forest, not back into $F_i$.

Two cases deserve mention. Since $C$ is unmerged, it is the vertex set
of a tree of $F_{i+1}$; hence $u'$ and $v'$ are disconnected in
$F_{i+1}$, and $r\ge i+2$ whenever $r$ exists. Moreover, $e'$ leaves
$C$ but may already be a forest edge of some tier $t\ge i+2$. Then
$r=t$, the $F_r$ path between $u'$ and $v'$ is the single edge $e'$, and
the rule selects $f=e'$; the exchange merely lowers the tier of $e'$
to $i+1$. Removing $f$ before inserting $e'$ ensures that no forest
ever holds two copies of an edge.

\begin{claim}
\label{clm:appd-top-path}
If $u'$ and $v'$ are connected in $F_D$, then $r$ equals the maximum tier
of an edge on their $F_D$ path, and this path is their path in $F_r$.
If they are disconnected in $F_D$, then $r$ does not exist.
\end{claim}
\begin{proof}
Let $t$ be the maximum tier on the $F_D$ path. All its edges lie in
$F_t$, so $u'$ and $v'$ are connected in $F_t$. If they were connected in
some $F_j$ with $j<t$, their $F_j$ path would also be their path in
$F_D\supseteq F_j$, so all its edges would have tier at most $j<t$, a
contradiction. Thus $r=t$, and the $F_r$ path is the $F_D$ path. The
last assertion follows from nestedness.
\end{proof}

\begin{lemma}[Effect of a repair]
\label{lem:appd-exchange}
Suppose the forests are nested and acyclic, $C$ is an unmerged tree of
$F_i$, and $e'=u'v'$ has exactly one endpoint in $C$, whether or not
$e'$ belongs to the current graph. Then the repair inserting $e'$
preserves nestedness and acyclicity and leaves $F_0,\ldots,F_i$
unchanged. On each layer $j>i$, the vertex partition into trees is
either preserved or coarsened.
\end{lemma}
\begin{proof}
Layers at most $i$ are not touched, since $r\ge i+2$. If $r$ exists,
then by minimality of $r$ the endpoints are disconnected in $F_{r-1}$,
so their $F_r$ path contains an edge of tier exactly $r$, and $f$ is
well defined. By nestedness, $f$ lies in $F_r,\ldots,F_D$, and the
$F_r$ path is the unique $u'$--$v'$ path in each of these forests.
Removing $f$ and adding $e'$ therefore preserves acyclicity and the
vertex partition on layers $r,\ldots,D$; this includes the case
$f=e'$. On layers $i+1,\ldots,r-1$, or on all layers above $i$ if $r$
does not exist, the endpoints are disconnected and adding $e'$ joins
two trees. Since $f\notin F_{r-1}$ and $e'$ is added to a contiguous
range of layers ending at $D$, nestedness is preserved.
\end{proof}

\subsection{The oracle process and seed conditioning}
\label{sec:appd-oracle}

Fix an oblivious update sequence, and let $G_s$ denote the current
graph after operation $s$. The \emph{ideal oracle process} uses the
exactly sound recovery rules above, and changes only the insertion
test: it links an inserted edge into every forest exactly when its
endpoints are disconnected in the true graph \emph{before} the
insertion. The oracle is an analysis device and need not be
efficient.

\begin{lemma}[Invariants of the oracle]
\label{lem:appd-invariants}
At every operation time $s$ of the ideal oracle process, including
between repairs, the forests $F_0\subseteq\cdots\subseteq F_D$ are
nested and acyclic, and every forest edge belongs to $G_s$.
\end{lemma}
\begin{proof}
Induct on operations and repairs. On an insertion of $uv$, the oracle
links $uv$ only if $u$ and $v$ are disconnected in $G_{s-1}$, hence in
every $F_j\subseteq G_{s-1}$; this creates no cycle. On a deletion,
the deleted edge is removed from every forest. A repair at layer $i$
queries an unmerged tree, so \Cref{lem:appd-exchange} applies;
the inserted edge belongs to $G_s$ by soundness of the ideal decoder,
and the exchange only removes the other edge $f$.
\end{proof}

Let $\omega_i$ contain the sampling seeds of the $q$ copies on layer
$i$, and let
\[
 \mathcal F_i=\sigma(\omega_0,\ldots,\omega_{i-1}).
\]
Fingerprint points are excluded from these seeds; the ideal process
does not use them. The remaining shared parameters are fixed in
preprocessing. The irreducible polynomial defining $\mathbb F_{2^b}$ is
deterministic. The prime $p$ of~\eqref{eq:appd-prime} is drawn
independently of all sampling seeds. Conditional on $p$, the
fingerprint points are drawn independently and uniformly from
$\mathbb F_p$, independently of the sampling seeds.

Since $p$ is independent of the sampling seeds, we may fix $p$ to an
arbitrary prime satisfying~\eqref{eq:appd-prime}; this leaves the
distribution of the seeds unchanged. All probabilities below are over
the sampling seeds and, where stated, the fingerprint points, for this
fixed $p$. Every bound holds for each such $p$, and hence also
unconditionally. The
ideal process does not depend on $p$. For every fixed operation time,
the true graph is deterministic because the update sequence is
oblivious.

\begin{lemma}[Projection onto the lower layers]
\label{lem:appd-projection}
At every fixed operation time in the ideal oracle process, including
between repairs, $F_0,\ldots,F_i$ are determined by the input prefix
and $\omega_0,\ldots,\omega_{i-1}$. In particular, $F_0,\ldots,F_i$,
the vertex sets of their trees, and which of these trees are active
are $\mathcal F_i$-measurable, while $\omega_i$ is independent of
$\mathcal F_i$.
\end{lemma}
\begin{proof}
Induct on updates and on the bottom-up repairs. Oracle insertions
and removal of the deleted edge depend only on the input. A repair
at layer $j\ge i$ changes no forest at or below $i$
(\Cref{lem:appd-exchange}). A repair at $j<i$ selects its queried
tree using $F_j$, its skip decision using $F_j$ and $F_{j+1}$ with
$j+1\le i$, and its recovered edge using $F_j$, the current graph,
and $\omega_j$.

If that edge's endpoints first become connected at a layer $r\le i$,
both $r$ and the edge $f$, which is selected by tier and edge
identifier, are determined by $F_0,\ldots,F_r$. If they are
disconnected through layer $i$, any cycle exchange occurs above the
projection; within it, the repair simply links the edge into layers
$j+1,\ldots,i$. Thus the projected change depends only on the stated
seeds. Activity of a tree depends additionally only on the
deterministic current graph.
\end{proof}

For a time $s$, a layer $i<D$, and a set $C\subseteq V$, define the
events
\begin{align*}
 \mathcal A^i_{s,C}&:=\{C\text{ is the vertex set of an active tree of }F_i
              \text{ at time }s\},\\
 \mathcal M^i_{s,C}&:=\{C\text{ is the vertex set of a tree of }F_i
              \text{ that is merged at layer }i+1\text{ at time }s\},
\end{align*}
evaluated after all repairs of operation $s$. By
\Cref{lem:appd-projection}, $\mathcal A^i_{s,C}$ is
$\mathcal F_i$-measurable.

\begin{lemma}[Conditional merge invariant]
\label{lem:appd-merge}
For every time $s$, layer $i<D$, and set $C\subseteq V$,
\[
 \Pr\bigl[\mathcal M^i_{s,C}\bigm|\mathcal F_i\bigr]\ge1-\varepsilon
 \qquad\text{almost surely on }\mathcal A^i_{s,C}.
\]
\end{lemma}
\begin{proof}
Fix $i$ and induct on $s$. Initially no tree is active. Consider
operation $s$, and assume the claim at time $s-1$ for all sets. We
partition $\mathcal A^i_{s,C}$ into $\mathcal F_i$-measurable cases. In each
case we exhibit an event $\mathcal E$ with $\Pr[\mathcal E\mid\mathcal F_i]\ge1-\varepsilon$
on that case and $\mathcal E\subseteq \mathcal M^i_{s,C}$ pointwise there, which proves
the claim. Throughout, we use that the partition of layer $i+1$ only
coarsens during the repairs of operation $s$
(\Cref{lem:appd-exchange}), so a tree of $F_i$ that is merged at
some moment of operation $s$, and is not itself changed afterwards,
stays merged.

\emph{Case 1: an insertion of $uv$ whose endpoints lie in one component
of $G_{s-1}$.} No forest changes. If $C$ were inactive at time $s-1$,
it would be an entire component of $G_{s-1}$, by
\Cref{lem:appd-invariants}; then $uv$ would not leave $C$, and $C$
would remain inactive. Hence $\mathcal A^i_{s-1,C}$ holds, and
$\mathcal E:=\mathcal M^i_{s-1,C}=\mathcal M^i_{s,C}$ works.

\emph{Case 2: an insertion of $uv$ joining two components of
$G_{s-1}$.} The oracle links $uv$ on every layer, so partitions only
coarsen. If $C$ contains neither endpoint, then $C$ is unchanged, its
outgoing edges are unchanged, and $\mathcal E:=\mathcal M^i_{s-1,C}$ works. Otherwise
$C=C_u\cup C_v$, where $C_u$ and $C_v$ are the trees of $F_i$
containing $u$ and $v$ at time $s-1$. If both were inactive, each was
an entire component of $G_{s-1}$, so $C$ is an entire component of
$G_s$, contradicting $\mathcal A^i_{s,C}$. Let $C^*$ be $C_u$ if it was active
and $C_v$ otherwise; this choice is $\mathcal F_i$-measurable. If
$C^*\subsetneq T^*$ for a tree $T^*$ of $F_{i+1}$ at time $s-1$, then
$T^*\setminus C^*$ lies in the old graph component of $C^*$, hence is
disjoint from $C$, while $T^*$ is contained in the tree of $F_{i+1}$
containing $C$ at time $s$. Thus $\mathcal E:=\mathcal M^i_{s-1,C^*}$ works.

\emph{Case 3: a deletion of $uv$, and $C$ contains neither endpoint.}
Deleting $uv$ splits only trees containing $uv$, and every resulting
tree contains $u$ or $v$. A repair at layer $j<i$ changes the vertex
partition of $F_i$ only by merging trees that contain the tree of
$F_j$ queried by the repair, which contains an endpoint; an exchange
preserves partitions. Hence $C$ was the vertex set of a tree of $F_i$
at time $s-1$, and it was active then because $G_s\subseteq G_{s-1}$.
Suppose $C\subsetneq T$ for a tree $T$ of $F_{i+1}$ at time $s-1$.
If deleting $uv$ splits $T$, the part containing $C$ also contains an
endpoint, which lies outside $C$; repairs then only coarsen layer
$i+1$. Hence $\mathcal E:=\mathcal M^i_{s-1,C}$ works.

\emph{Case 4: a deletion of $uv$, and $C$ contains an endpoint, say
the first one processed among those in $C$.} When layer $i$ is
processed, its partition is already final for this operation: repairs
at layers $j<i$ have finished, and repairs at layers $j\ge i$ do not
change $F_i$. So the algorithm considers exactly $C$ at layer $i$. Let
$\mathcal E:=\mathcal R^i_{s,C}$ be the event that the hypothetical amplified ideal
recovery with seed $\omega_i$ on the cut $\delta_{G_s}(C)$ succeeds,
whether or not the algorithm executes that call. For fixed $C$, this
event depends only on $\omega_i$, which is independent of
$\mathcal F_i$; since $\delta_{G_s}(C)\ne\emptyset$ on $\mathcal A^i_{s,C}$,
\eqref{eq:appd-epsilon} gives $\Pr[\mathcal E\mid\mathcal F_i]\ge1-\varepsilon$
there. If the call is skipped, $C$ is already merged; if it is
executed, it returns the same edge as the hypothetical call, and the
repair makes $C$ merged. Later repairs at layer $i$ only coarsen
layer $i+1$. Hence $\mathcal E\subseteq \mathcal M^i_{s,C}$ pointwise on this case.

We never condition on a skip decision or on earlier answers of the
layer-$i$ copies.
\end{proof}

\subsection{The deletion-sensitive component bound}
\label{sec:appd-bottom}

Let $H_s$ be the historical graph containing all edge labels ever
inserted by operation time $s$. Unlike the backbone of \Cref{subsec:rand-upper-backbone}, the
forest $F_0$ maintained here is not a subforest of a fixed spanning
forest of $H_s$, since edges can leave $F_0$ and later re-enter it.
The following argument therefore counts components instead.

\begin{lemma}[At most $2k$ active bottom-layer trees]
\label{lem:appd-bottom}
At every operation time $s$, the ideal oracle's forest $F_0$ has at
most $2k$ active trees.
\end{lemma}
\begin{proof}
Write $\operatorname{cc}(J)$ for the number of connected components
of $J$, and put
\[
 d_s=\operatorname{cc}(F_0)-\operatorname{cc}(H_s)\ge0,
\]
where nonnegativity uses $F_0\subseteq G_s\subseteq H_s$
(\Cref{lem:appd-invariants}). Initially $d_0=0$. If an insertion
joins distinct historical components, it also joins distinct current
graph components. The oracle links it into $F_0$, so both component
counts decrease by one. Any other insertion leaves
$\operatorname{cc}(H_s)$ unchanged and cannot increase
$\operatorname{cc}(F_0)$. A deletion leaves $H_s$ unchanged and
increases $\operatorname{cc}(F_0)$ by at most one. Repairs do not
change $F_0$. Hence $d_s$ is at most the number of deletions so far,
and in particular $d_s\le k$. Each deletion is counted, even if its
edge identifier was deleted before and subsequently reinserted.

For each historical component $Q$, let $n_Q$ be the number of $F_0$
trees contained in it. Then $\sum_Q(n_Q-1)=d_s$. If $n_Q=1$, that
tree is all of $Q$ and is inactive, since no historical edge leaves
$Q$. Therefore, if $X_0$ is the number of active $F_0$ trees,
\[
 X_0\le\sum_{Q:n_Q>1}n_Q
       \le2\sum_Q(n_Q-1)
       =2d_s\le2k.
\]
\end{proof}

Since also $d_s\le\operatorname{cc}(F_0)-1\le n-1$, the bound improves
to $X_0\le2\min\{k,n-1\}$. We do not need this refinement.

\subsection{Progress and simultaneous correctness}
\label{sec:appd-progress}

Fix an operation time in the ideal oracle process. Let $X_i$ be the
number of active $F_i$ trees, and let $Y_i$ count those that are
unmerged at layer $i+1$. Both are functions of the final states at
this fixed time. The nested partitions give the deterministic bound
\begin{equation}
 X_{i+1}\le\frac{X_i+Y_i}{2}\le X_i.
 \label{eq:appd-contraction}
\end{equation}
To see this, note that an inactive tree of $F_i$ is an entire
component of the current graph; since $F_{i+1}$ is contained in the
current graph (\Cref{lem:appd-invariants}), it is also a tree of
$F_{i+1}$, and an inactive one. Hence every active tree of $F_{i+1}$
is a union of active trees of $F_i$. At most $Y_i$ of them consist of
a single $F_i$ tree, namely an unmerged active one, and every other
one contains at least two active $F_i$ trees. Thus
$2X_{i+1}-Y_i\le X_i$.

Call layer $i$ \emph{good} if $Y_i\le X_i/4$, including when $X_i=0$.
On a good layer, \eqref{eq:appd-contraction} gives
$X_{i+1}\le(5/8)X_i$. Since $Y_i=\sum_C\mathbf 1[\mathcal A^i_{s,C}]
(1-\mathbf 1[\mathcal M^i_{s,C}])$ at the fixed time $s$,
\Cref{lem:appd-merge} and linearity of expectation give
\[
 \mathbb E[Y_i\mid\mathcal F_i]\le\varepsilon X_i.
\]
Since $X_i$ is $\mathcal F_i$-measurable, Markov's inequality yields
\begin{equation}
 \Pr[\text{layer }i\text{ is bad}\mid\mathcal F_i]
 \le4\varepsilon.
 \label{eq:appd-bad-layer}
\end{equation}
No independence between trees on the same layer is used.

Choose the integer parameters
\begin{align}
 D&=2\bigl(\lceil\log_{8/5}(2k)\rceil+1\bigr),
 \label{eq:appd-depth}\\
 q&=\left\lceil
       \frac{4+(2/D)\log_2(2k/\delta)}{\log_2(8/7)}
     \right\rceil.
 \label{eq:appd-q}
\end{align}
To see the dependence of $q$ on the confidence parameter, expand
\[
 \log_2(2k/\delta)=\log_2(2k)+\log_2(1/\delta).
\]
Since $D=\Theta(\log(1+k))$ and
$\log_2(2k)=\Theta(\log(1+k))$, the first term divided by $D$
is bounded by constants. The additive constant in
\eqref{eq:appd-q} and the ceiling ensure a constant baseline number
of copies. Thus the remaining confidence cost is
$\Theta(\L(\delta)/D)$. In particular,
\begin{equation}
 D=\Theta(\log(1+k)),\qquad
 q=\Theta\left(1+\frac{\L(\delta)}{\log(1+k)}\right),\qquad
 Dq=\Theta\bigl(\log(1+k)+\L(\delta)\bigr).
 \label{eq:appd-parameters}
\end{equation}

\begin{lemma}[Failure at a fixed deletion time]
\label{lem:appd-fixed-time}
At every fixed deletion time, the ideal oracle's $F_D$ fails to be a
spanning forest of the current graph with probability at most
$\delta/(2k)$.
\end{lemma}
\begin{proof}
For bad-layer events $\mathcal B_i$, each $\mathcal B_i$ is
$\mathcal F_{i+1}$-measurable. If $i_1<\cdots<i_j$, iterated
conditioning in \eqref{eq:appd-bad-layer} gives
\[
 \Pr[\mathcal B_{i_1}\cap\cdots\cap\mathcal B_{i_j}]\le(4\varepsilon)^j.
\]
This does not assert independence of the events.

By \Cref{lem:appd-bottom}, $X_0\le2k$. If at least $D/2$ of the
$D$ layers are good, monotonicity and contraction give
$X_D\le2k(5/8)^{D/2}\le\tfrac58<1$, since
$D/2\ge\log_{8/5}(2k)+1$; hence $X_D=0$. Then every tree of $F_D$ has no outgoing edge in the
current graph and is therefore an entire component; since $F_D$ is a
forest contained in the current graph (\Cref{lem:appd-invariants}), it
is a spanning forest. Failure
therefore implies that fewer than $D/2$ layers are good, so more than
$D/2$ layers are bad and, as $D$ is even, all layers of some set of
exactly $D/2$ layers are bad. A union bound over these sets gives
\[
 \Pr[X_D>0]
 \le2^D(4\varepsilon)^{D/2}
 =2^{2D}(7/8)^{qD/2}
 \le\frac{\delta}{2k}.
\]
Taking logarithms, the last inequality is equivalent to
\[
 \log_2(8/7)\cdot qD/2\ge2D+\log_2(2k/\delta),
\]
which holds by the definition~\eqref{eq:appd-q} of $q$.
\end{proof}

Amplification is therefore required to make sufficiently many layers
good, rather than to make every individual recovery call succeed. This
is why the confidence cost in the number of copies per layer is
divided by $D$.

An insertion preserves a correct oracle top forest: it links two
trees exactly when the insertion joins distinct true graph components.
The top forest is also correct initially. Thus a first failure can
occur only after a deletion. Since the sequence has at most $k$
deletion operations, a union bound over these fixed times gives
\begin{equation}
 \Pr[\text{the ideal oracle's top forest is ever incorrect}]
 \le\delta/2.
 \label{eq:appd-global-progress}
\end{equation}
Repeated connectivity queries that inspect $F_D$ create no new failure
events. In particular, no assumption on the number of such queries is
needed, provided that the updates remain oblivious.

\subsection{Fingerprint errors and coupling}
\label{sec:appd-soundness}

We use the singleton decoder of \Cref{subsec:rand-upper-recovery}, with a prime $p>2N$.
For a sampled signed cut vector $z\in\{-1,0,1\}^N$, it maintains
\[
 m_1(z)=\sum_e\operatorname{id}(e)z_e,\qquad
 m_2^\alpha(z)=\sum_e\alpha^{\operatorname{id}(e)}z_e
 \quad\text{in }\mathbb F_p.
\]
Every true singleton is accepted for every fingerprint point
$\alpha$. For a nonsingleton vector, the probability of acceptance
over an independent uniform $\alpha\in\mathbb F_p$ is at most $N/p$
by \Cref{lem:rand-upper-decoder}. The polynomial defining acceptance is nonzero: the
signed coefficients do not vanish modulo $p>2N$, and the degree
is at most $N$.

At each deletion, the ideal oracle invokes at most $2D$ recoveries.
For every invoked recovery, include all $qb$ singleton tests on
its queried cut, even tests skipped after the first successful copy or
level. These \emph{potential tests} number at most
\[
 M_{\max}=2kDqb.
\]
Choose a prime satisfying
\begin{equation}
 P<p\le2P,\qquad
 P=\max\{2N,\,2NM_{\max}/\delta\}.
 \label{eq:appd-prime}
\end{equation}
Bertrand's postulate ensures existence. Since $k\le n^2$ and
$\delta\ge n^{-\kappa}$, \eqref{eq:appd-parameters} gives
$Dq=O_\kappa(\log n)$ and $M_{\max}=O_\kappa(n^2\log^2 n)$; hence
$\log p=O_\kappa(\log n)$. Field elements and arithmetic therefore use
a constant number of word-RAM words, with constants depending on
$\kappa$.

Recall that $p$ is fixed (\Cref{sec:appd-oracle}). Condition further on
all sampling seeds. This fixes the entire ideal oracle execution and
all vectors in its potential tests, while the fingerprint points remain
independent and uniform in $\mathbb F_p$.
A union bound,
without requiring independence between tests using the same point,
gives
\begin{equation}
 \Pr[\text{some potential nonsingleton test is accepted}]
 \le M_{\max}N/p<\delta/2.
 \label{eq:appd-fingerprint}
\end{equation}
We bound this event without first conditioning on successful progress
of the oracle.

Run the actual algorithm and the ideal oracle with the same sampling
seeds and fingerprint points. On the complement of the events in
\eqref{eq:appd-global-progress} and \eqref{eq:appd-fingerprint}, they
have identical logical forest states throughout the sequence. Induct
on updates. Before an insertion, their top forests agree and are
correct, so the actual top-forest connectivity test equals the
oracle's true pre-insertion connectivity test. On a deletion, their
initial removals agree. For each subsequent recovery, they encounter
the same queried cut and make the same skip decision. Every singleton
test accepts; every nonsingleton potential test
rejects. They therefore return the same first candidate and perform
the same deterministic repair. This proves agreement by induction over
the repair calls and updates.

Consequently, with probability at least $1-\delta$, the actual top
forest is a spanning forest after every update. This argument does
not condition a reused sampling seed on its previous answers or on
the absence of earlier failures. Without an exact current-edge
membership dictionary, we make no unconditional one-sided-error claim.

\subsection{Worst-case running time and space}
\label{sec:appd-complexity}

\subparagraph*{Data structures.}
In every recovery copy, represent the forest by Euler-tour trees stored
in balanced search trees with worst-case $O(\log n)$ split and join,
such as red-black trees; treaps or splay trees would only give expected
or amortized bounds. In each Euler tour, every vertex has exactly one
designated occurrence, which carries the vertex's sketch and weight
one; all other occurrences carry zero sketches and weight zero. If a
cut removes the designated occurrence of a vertex, another occurrence
of that vertex becomes designated, at the cost of $O(\log n)$ aggregate
updates. Each node stores the sum of the $O(b)$ field elements and the
total weight in its subtree. Hence the root of an Euler-tour tree holds
the summed sketch of its tree and the number of its vertices, which
supports recovery, the size comparison of the skip test, and
connectivity tests on layers $0,\ldots,D-1$.

Represent $F_D$ by top trees with the worst-case implementation of
Alstrup, Holm, de Lichtenberg, and Thorup~\cite{TopTrees05}, based on
topology trees, which supports link, cut, connectivity, and
path-maximum queries in $O(\log n)$ worst-case time; an amortized
implementation, for example via splay-based link-cut trees, would not
establish the stated bounds. Edge weights in this structure are pairs
$(\text{tier},\operatorname{id})$ compared by tier and then by reversed
identifier, so that a path-maximum query returns the edge $f$ of the
cycle-exchange rule. Finally, to answer connectivity queries faster, we
also store the Euler tour of $F_D$ in balanced $\Theta(\log n)$-ary
trees, as in~\cite{HdLT01}, with the same designated-occurrence weights.
Their roots give the tree sizes of $F_D$, which the skip test at layer
$D-1$ compares with those of $F_{D-1}$. These trees have height
$O(\log n/\log\log n)$, and a link or cut of $F_D$ takes
$O(\log^2 n/\log\log n)$ worst-case time. Every update changes $F_D$
at most $O(D)$ times, so this copy adds $O(D\log^2 n/\log\log n)$
worst-case time per update, which the bounds below absorb.

\subparagraph*{Update time.}
An insertion updates the sketches of $Dq$ copies, tests connectivity
in $F_D$, and performs at most $D+1$ forest links, propagated to the
$q$ copies on each affected layer. Its cost is $O_\kappa(Dq\log^2 n)$.

A deletion first removes the edge from the sketches of all copies and
from every forest containing it, in $O_\kappa(Dq\log^2 n)$ time. It then
performs at most $2D$ skip tests and recoveries, each costing
$O_\kappa(q\log^2n)$. For a successful recovery, a single connectivity test
and path-maximum query in $F_D$ determine $r$ and $f$ by
Claim~\ref{clm:appd-top-path}, in $O(\log n)$ time; no scan over the
layers is needed. The repair then changes at most $D$ forest layers,
with at most one cut and one link on each changed layer, propagated to
the $q$ copies of the layer, and updates the tier of $e'$ in $F_D$.
A repair therefore costs $O_\kappa(Dq\log^2 n)$, and the entire deletion
costs $O_\kappa(D^2q\log^2 n)$ worst-case time. This operation count also
holds on unsuccessful executions: the endpoint check and
\Cref{lem:appd-exchange}, which does not require the candidate to
be a current edge, keep every forest nested and acyclic even if a false
candidate is accepted.

Substituting \eqref{eq:appd-parameters}, and setting $L:=Dq$, yields
\begin{align*}
 T_{\mathrm{ins}}
  &=O_\kappa(Dq\log^2 n)=O_\kappa(L\log^2 n),\\
 T_{\mathrm{del}}
  &=O_\kappa(D^2q\log^2 n)
    =O_\kappa(L\log^2 n\log(1+k)).
\end{align*}
This proves \Cref{thm:appd-dynamic}. A connectivity query compares
the roots of the trees containing the two query vertices, in
$O(\log n/\log\log n)$ worst-case time.

\subparagraph*{Preprocessing.}
Before the first update, the data structure first selects the prime
in~\eqref{eq:appd-prime}, using randomness independent of all other
choices: it samples uniform integers in $(P,2P]$ and tests each with
the deterministic polynomial-time AKS primality test~\cite{AKS04}. By the prime number theorem,
$O(\log P)=O_\kappa(\log n)$ samples suffice in expectation, so this
takes expected $\operatorname{poly}(\log n)$ time. The irreducible
polynomial defining $\mathbb F_{2^b}$ is computed deterministically in
$\operatorname{poly}(\log n)$ time~\cite{Shoup90}. The data structure
then samples all sampling seeds and, given $p$, all fingerprint points,
and builds $Dq$ copies over $n$ singleton trees with zero sketches, in
$O(nDqb)=O(nL\log n)$ time. These costs are incurred once and are not
charged to updates.

\subparagraph*{Space.}
For completeness, this direct full-scale implementation uses
\begin{equation}
 O(nDq\log^2 n)=O(nL\log^2 n)\text{ bits},
 \label{eq:appd-space}
\end{equation}
including stored seeds, forest representations, tier metadata, and
reused working memory. This is not a claim that the dynamic structure
achieves the optimal streaming space bounds of \Cref{thm:main}. The
truncated samplers and historical-forest scans in \Cref{subsec:rand-upper-recovery} require
additional implementation arguments before their space bounds can
be combined with worst-case polylogarithmic recovery time.

\subparagraph*{Comparison.}
For constant $\delta$, the above bounds become
$O(\log^2 n\log(1+k))$ per insertion and
$O(\log^2 n\log^2(1+k))$ per deletion. For $\delta=n^{-c}$ with
fixed $c>0$, they become $O_c(\log^3 n)$ per insertion and
$O_c(\log^3 n\log(1+k))$ per deletion. Wang~\cite{Wang15} obtains
$O(\log^3 n)$ worst-case time per insertion and $O(\log^4 n)$ per
deletion with failure probability $n^{-c}$, improving the
$O(\log^4 n)$ and $O(\log^5 n)$ bounds of Kapron, King, and
Mountjoy~\cite{KKM13}. Our deletion bound is therefore smaller by a
factor $\Theta(\log n/\log(1+k))$, which is $\omega(1)$ when
$\log(1+k)=o(\log n)$, and it matches Wang's bound when
$k=n^{\Theta(1)}$. Our connectivity queries match Wang's
$O(\log n/\log\log n)$ worst-case time.
By a different approach, Meierhans and Probst Gutenberg~\cite{MG26}
obtain a Las Vegas algorithm with expected $\widetilde O(1)$ worst-case
update time, and Meierhans, Probst Gutenberg, and Yeh~\cite{MPGY26}
obtain $\widetilde O(1)$ worst-case update time with high probability;
both work against an adaptive adversary. The comparison above is within
the cutset-based line of work.
\section{Randomized Upper Bound for Variable Failure Probability}
\label{sec:high-prob-upper}
This section is dedicated to the proof of \cref{thm:high-prob-upper}. We remind readers that $K=\min\set{k,\binom{n}{2}}$.
\begin{theorem}
\label{thm:high-prob-upper}
Fix \(c>0\). Let \(n\ge 2\), \(k\ge 0\), and \(n^{-c}\le\delta\le 1/3\). Set \(B=\lceil\log((K+1)/\delta)\rceil\).
There is a randomized one-pass streaming algorithm that outputs a spanning forest of the final graph with probability at least \(1-\delta\) using
\[
    O_c\!\left(nB\log n\,\log\!\left(1+\frac{K+1}{B}\right)\right)
\]
bits of space.
\end{theorem}
Together with \cref{thm:general_LB}, this theorem yields the second part of \cref{thm:main}.

\subsection{General Algorithm}
\label{subsec:high-prob-algorithm}
\begin{itemize}
    \item When \(K\le B\), maintain \(K+1\) edge-disjoint historical forests~\cite[Lemma~2.1]{nagamochi1992linear}. On insertion, assign an unstored edge to the first forest in which it joins two components, if one exists. Keep each assigned edge in its forest and maintain its presence bit as in \cref{subsec:rand-upper-backbone}. Return a spanning forest of the stored edges present at the end of the stream.
    \item When \(K>B\), run the contraction of \cref{subsec:rand-upper-overview} from \(F_0\) for \(R=2(B+2)\) rounds. Each round uses an independent recovery bundle from \cref{subsec:rand-upper-recovery} with \(q_s=40\) singleton-sampling copies and \(q_m=16\) historical forests. Each copy draws its hash with \(L=\ell_1\) and stores only the \(J=\ell_1-\ell_0+1\) levels from \(\ell_0=\lfloor\log B\rfloor\) to \(\ell_1=\lceil\log(2K)\rceil\).
    All rounds share one additional sparse-recovery sketch per vertex (\cref{subsec:high-prob-sparse}). For each component, first query its aggregated sketch. If the decoder returns a vector, use the smallest-ID edge in its support as the candidate. Otherwise, query the current round's bundle.
\end{itemize}

In the branch \(K\le B\), consider a final edge \(uv\) that was never stored. At its first insertion, every forest already contained a \(u\)-to-\(v\) path. These \(K+1\) paths are edge-disjoint. At most \(K\) historical edges are absent by~\eqref{eq:rand-upper-missing}, so at least one path survives. Thus the present stored edges connect the endpoints of every final edge. The forests use \(O(n(K+1)\log n)\) bits. Since \((K+1)/B\le 2\) and \(\log(1+x)=\Theta(x)\) for \(0\le x\le 2\), we have \(B\log(1+(K+1)/B)=\Theta(K+1)\), including when \(k=0\).

For the remainder, assume \(K>B\). Each round has at most \(n\) components. Counting one shared-sketch test and up to \(q_sJ\) singleton tests per component gives at most \(Q_{\max}\coloneqq nR(1+q_sJ)\) potential tests overall. By Bertrand's postulate, choose a prime \(2Q_{\max}N/\delta<p\le4Q_{\max}N/\delta\). Then \(p>2N\) for singleton decoding and \(Q_{\max}N/p<\delta/2\) for fingerprint verification. Since \(K\le N\) and \(\delta\ge n^{-c}\), we have \(B,R=O_c(\log n)\), \(J=O(\log n)\), and \(\log p=O_c(\log n)\). Draw and store all random choices before the stream.

\subsection{The shared sparse-recovery sketch}
\label{subsec:high-prob-sparse}

We use the standard Vandermonde sketch for sparse recovery
(see, e.g.,~\cite[Appendix~E.1]{feng_et_al:ICML.2024}).
For \(x\in\{-1,0,1\}^N\), define its \(2B\) \emph{moments}
\[
    a_j(x)\coloneqq\sum_{e\in\mathcal E}\id(e)^j x_e
    \quad\text{in }\mathbb F_p,
    \qquad j=0,\ldots,2B-1.
\]
Each vertex \(v\) stores these moments of its signed incidence vector
\(x^{(v)}\) and the fingerprint \(m_2^{\alpha_0}(x^{(v)})\) from
\cref{subsec:rand-upper-recovery}. The point \(\alpha_0\) is uniform in
\(\mathbb F_p\), independent of all bundles, and shared across vertices
and rounds. Summing the sketches over a component \(C\) gives the sketch
of its signed cut vector by~\eqref{eq:rand-upper-cut-cancellation}.

Given the sketch of \(x\), the decoder uses the moments to find a vector
\(z\in\{-1,0,1\}^N\) of support size at most \(B\) with the same moments.
It returns \(z\) if such a vector exists, is nonzero, and satisfies
\(m_2^{\alpha_0}(z)=m_2^{\alpha_0}(x)\). Otherwise it rejects.

\begin{lemma}
\label{lem:high-prob-sparse}
For every fixed \(x\in\{-1,0,1\}^N\), the decoder recovers \(x\)
for every \(\alpha_0\) if \(1\le|\supp(x)|\le B\), and rejects if
\(x=0\).
If \(|\supp(x)|>B\), it accepts with probability at most \(N/p\).
It uses \(O(B\log p)\) bits of workspace.
\end{lemma}

\begin{proof}
Suppose two distinct vectors \(z,z'\), each with support size at most
\(B\), have the same moments.
Their difference has support size \(t\le2B\).
The first \(t\) moment equations form a square Vandermonde system
on these \(t\) distinct edge IDs.
Since \(p>N\), the system is invertible, so \(z=z'\) over
\(\mathbb F_p\), and hence over the integers because \(p>2\).
This proves uniqueness and the assertions for support size at most
\(B\).

If \(|\supp(x)|>B\), any candidate \(z\) differs from \(x\) and is
determined independently of \(\alpha_0\).
Acceptance implies that \(\alpha_0\) is a root of the nonzero polynomial
\(\sum_{e\in\mathcal E}(x_e-z_e)X^{\id(e)}\in\mathbb F_p[X]\).
Its degree is at most \(N\), giving the claimed error bound.

The decoder enumerates one signed support of size at most \(B\) at a time, storing its edge IDs, signs, and \(2B\) moments in \(O(B\log p)\) bits. It evaluates field powers by repeated squaring within the same workspace.
\end{proof}

\subsection{Correctness and space}
\label{subsec:high-prob-analysis}

Fix an oblivious stream. Use the ideal process of
\cref{subsec:rand-upper-contraction}, with exact singleton tests.
Its shared decoder returns the true cut vector for cut sizes \(1\) through
\(B\) and rejects otherwise. Thus the hashes and ranks of the preceding
rounds determine the current partition, independently of the current bundle.

Condition on that history and consider an active component with cut size
\(s\). If \(s\le B\), the ideal shared decoder succeeds. If \(B<s\le K\),
the stored range contains \(\ell=\lceil\log(2s)\rceil\), the level used in
\cref{lem:rand-upper-recovery} to isolate an edge. For \(s>K\), the
historical forests give the guarantee in the same lemma. With the stated
numbers of copies and forests, the conditional probability of obtaining
no candidate is at most \(1/64\) in every case.

Let \(A_r\) count the active components after \(r-1\) rounds.
Every ideal candidate is a valid outgoing edge, and \(A_1\le2K\)
by \cref{lem:rand-upper-backbone}. The contraction
estimate in the proof of \cref{lem:rand-upper-convergence} therefore gives \(\Pr[A_{R+1}>0]\le2K(65/128)^R\le 2K\,2^{-(B+2)}\le \delta/2\),
since \((65/128)^2<1/2\) and \(2^B\ge(K+1)/\delta\).
When \(A_{R+1}=0\), the ideal forest spans \(G\) by \cref{lem:rand-upper-backbone}.

Now fix all sampling hashes and edge ranks. This fixes every query vector
along the ideal execution independently of every fingerprint point. By
\cref{lem:high-prob-sparse,lem:rand-upper-decoder}, each test on these
vectors disagrees with its ideal answer with probability at most \(N/p\).
A union bound gives probability at most \(Q_{\max}N/p<\delta/2\) of
any disagreement, even with the same \(\alpha_0\) in every round.
If all these tests agree, both executions start from \(F_0\) and choose
the same candidates on each successive partition, as in the coupling
proof of \cref{lem:rand-upper-coupling}. Adding the contraction and
fingerprint error bounds gives failure probability at most \(\delta\).

The shared sketches use \(O(nB\log p)\) bits, with \(O(B\log p)\) reusable
decoding workspace. Storing \(J\) levels per copy in the bundle accounting
of \cref{lem:rand-upper-bundle-space} gives \(O(nR(J\log p+\log n))\)
bits for all bundles. The backbone,
component data, output forest, and candidates take \(O(n\log n)\) bits.
The remaining workspace fits within these bounds as in
\cref{subsec:rand-upper-completion}.
For \(K>B\), the number of stored levels is
\[
    J=\lceil\log(2K)\rceil-\lfloor\log B\rfloor+1
      =\log(K/B)+O(1).
\]
Thus \(J+1=O(\log(1+(K+1)/B))\). Using \(R=O(B)\) and \(\log p=O_c(\log n)\), the total space is
\[
    O_c\bigl(n\log n\, [B+R(J+1)]\bigr)
    = O_c\!\left(nB\log n\,\log\!\left(1+\frac{K+1}{B}\right)\right).
\]
Together with the deterministic branch for \(K\le B\), this proves
\cref{thm:high-prob-upper}.

\paragraph*{Matching bounds.}
For \(1\le k\le n^2\), \(K=\Theta(k)\) and \(B=\Theta(t)\) with \(t=\log(k/\delta)\), giving the upper bound in \cref{thm:main}.
Apply \cref{thm:general_LB} on \(2u\) vertices, where \(u=\lfloor n/2\rfloor\), attaching a leaf if \(n\) is odd.
If \(k\le u\), then \(K'=k\) and \(\Gamma=\min\{k,t\}=O_c(\log n)\), so \(\log(u/\Gamma)=\Theta_c(\log n)\).
By the convention of \cref{sec:preliminaries}, the remaining factor is \(\Gamma\log(1+k/\Gamma)=\Theta(t\log(1+k/t))\), since the two expressions agree for \(k>t\) and are both \(\Theta(k)\) for \(k\le t\).
If \(u<k\le n^2\), then \(K'=u\) and \(\Gamma,t=\Theta_c(\log n)\), giving \(\Theta_c(n\log^3 n)\) in both bounds.

\end{document}